\documentclass[%
 aip,
 amsmath,amssymb,
 reprint,%
]{revtex4-1}
\usepackage{float}
\usepackage{graphicx}
\usepackage{dcolumn}
\usepackage{bm}
\usepackage{gensymb}

\usepackage[utf8]{inputenc}
\usepackage[T1]{fontenc}
\usepackage{mathptmx}
\usepackage{color}
\usepackage{hyperref}
\newcommand{\new}[1]{{\color{blue}{#1}}}

\newcommand{\old}[1]{}
\usepackage{float} 

\begin{document}

\preprint{AIP/123-QED}

\title[\textbf{Published on: Review of Scientific Instruments 93, 043704 (2022); https://doi.org/10.1063/5.0078423}]{{
\new{Design and implementation of a device based on an off-axis parabolic mirror to perform luminescence experiments in a Scanning Tunneling Microscope}}}

\author{Ricardo Javier Peña Román} \affiliation{"Gleb Wataghin" Institute of Physics, University of Campinas – UNICAMP, 13083-859, Campinas, SP, Brazil.}

\author{Yves Auad}
\author{Lucas Grasso}
\author{Lázaro Padilha}
\author{Fernando Alvarez}\affiliation{"Gleb Wataghin" Institute of Physics, University of Campinas – UNICAMP, 13083-859, Campinas, SP, Brazil.}

\author{Ingrid David Barcelos}\affiliation{ Brazilian Synchrotron Light Laboratory (LNLS), Brazilian Center for Research in Energy and Materials (CNPEM), 13083-970, Campinas, SP,
Brazil.}

\author{Mathieu Kociak}\affiliation{Université Paris-Saclay, CNRS, Laboratoire de Physique des Solides, 91405, Orsay, France}

\author{Luiz Fernando Zagonel }\email{The author to whom correspondence may be addressed: zagonel@ifi.unicamp.br.} \affiliation{"Gleb Wataghin" Institute of Physics, University of Campinas – UNICAMP, 13083-859, Campinas, SP, Brazil.}


\begin{abstract}
We present the design, the implementation, and illustrative results of a light collection/injection strategy based on an off-axis parabolic mirror collector for a low-temperature Scanning Tunneling Microscope (STM). This device allows to perform STM induced Light Emission (STM-LE) and Cathodoluminescence (STM-CL) experiments as well as \emph{in situ} Photoluminescence (PL) and Raman spectroscopy as complementary techniques. Considering the \emph{Étendue} conservation and using an off-axis parabolic mirror, it is possible to design a light collection and injection system that displays 72$\%$ of collection efficiency (considering the hemisphere above the sample surface) while maintaining high spectral resolution and minimizing signal loss. The performance of the STM is tested by atomically resolved images and scanning tunneling spectroscopy results on standard sample surfaces. The capabilities of our system are demonstrated by performing STM-LE on metallic surfaces and 2D semiconducting samples, observing both plasmonic and excitonic emissions. Also, we carried out \emph{in situ} PL measurements on semiconducting monolayers and quantum dots, and \emph{in situ} Raman on graphite and hexagonal boron nitride (h-BN) samples. Additionally, STM-CL and PL were obtained on monolayer h-BN gathering luminescence spectra that are typically associated with intragap states related to carbon defects. The results show that the flexible and efficient light injection and collection device based on an off-axis parabolic mirror is a powerful tool to study several types of nanostructures with multiple spectroscopic techniques in correlation with their morphology at the atomic scale and electronic structure.

\end{abstract}

\maketitle

\section{\label{sec1} Introduction}

In the past decades, quantum-confined structures, plasmonic nanostructures, and other optically active systems which rely on nanoscopic details of their morphology or electronic structure have been extensively created, studied, and applied in several fields. \cite{Ogletree2015} This is particularly true for III-V, III-N, II-VI quantum structures which can form Quantum Dots (QDots) and Nanowires (NW) of great interest in nanotechnology.\cite{Jacopin2011,Ruhle2010,Agarwal2008,Zhang2015}. More recently, monolayers of some two-dimensional (2D) materials like the transition metal dichalcogenides (TMDs) are of great technological interest due to their direct band gap.\cite{Mak2010,Xu2013,Zheng2018,Samadi2018} Moreover, the quantum efficiency (QE) of some 2D materials has been observed to depend on the defect density.\cite{Tonndorf2015,Wei2014,Tongay2013,Wu2017} In general, objects of interest in nanotechnology are smaller than the optical diffraction limit and so are the relevant details in their structure, including point defects which are in the atomic scale. Widespread spectroscopic investigations by optical absorption or photoluminescence (PL) are hence usually limited to ensembles of these materials and cannot resolve some relevant features of their light emission and morphology.

Therefore, techniques with spatial resolution in the nanometric or atomic scales and spectroscopic capabilities in the meV are thus necessary for many problems in the field. ~\cite{Ogletree2015,Gustafsson1998,KOCIAK2014} Two analog but completely different approaches are known to reach atomic resolution and provide spectroscopy data. On the one hand, using fast electrons, (Scanning) Transmission Electron microscopes ((S)TEM) provide atomic resolution imaging with several contrast methods and also spectroscopic techniques like Electron Energy Loss Spectroscopy (EELS). In the IR to UV range, EELS gives access to several physical properties such as dielectric function and band gap, as well as to different physical excitation including plasmons, excitons, and phonons.\cite{KOCIAK2014} On the other hand, using a physical probe with low energy electrons, Scanning Tunneling Microscopy (STM) also reaches atomic resolution and provides a spectroscopic technique to measure the local densities of states (LDOS) in the conduction and valence bands by means of Scanning Tunneling Spectroscopy (STS).\cite{wiesendanger_1994,Plumadore} Moreover, beyond electronic spectroscopic techniques, both STEM and STM can be associated with light collection devices to perform cathodoluminescence (STEM-CL) and STM Induced Light Emission (STM-LE), respectively. Recent reviews have been published about these two techniques highlighting, in several physical systems, the great potential and interest of locally and electrically excited luminescence.\cite{KOCIAK2017,Kuhnke2017} It is important to mention that CL can also be performed similarly in Scanning Electron Microscopes (SEMs) with similar interests as in STEMs and was the subject of a recent review as well.\cite{CoenenCentury} In particular, and within the scope of the present work, STM-LE has already yielded great insights for plasmonics, oxides, molecules, III-V and II-VI semiconductors among others \cite{Stavale2013,Stavale2014,Moal2013,Lutz2011,Chen2010,Qiu2003} and more recently in the 2D semiconductors MoSe$_2$ and WSe$_2$.\cite{Pommier2019,pechou2020,Pena2020,DOAMARAL2021} 

One key issue in STM-LE studies is the light collection strategy since the light emission yield is about 10$^{-4}$ to 10$^{-6}$ photons per electron \cite{Sakurai2005}. Such quantum yield is much smaller than what is achieved by STEM-CL for instance, which can be as high as 1 photon per electron \cite{KOCIAK2017,Zagonel2016}. Nevertheless, the collection efficiency is optimized for STEM-CL to provide sufficient spectral signal-to-noise ratio and acquisition speed. In STM-LE, given the tunneling current typically used in STM, typically from 10 pA to 10 nA, one would not expect much more than about 10$^{7}$ photons per second (p/s) from a tunneling junction, and as few as 10$^{1}$ p/s might be extracted in some cases. Knowing that, it is no surprise if the light collection efficiency and transmission efficiency is even more critical in STM-LE than in STEM-CL. 

 Integrating efficient light collection devices in an STM is not easy and therefore several strategies have been considered and some STMs were constructed specifically to allow STM-LE studies.\cite{Berndt1991} In a recent paper by K. Edelmann et al. \cite{Edelmann2018}, for instance, the STM tip itself was integrated inside a light collecting mirror in a Low-Temperature and Ultra High Vacuum STM (LT-UHV STM), reaching 75$\%$ of collection efficiency. Using another high collection efficiency device, E. Le Moal et al. employed an STM mounted on an inverted optical microscope using a high numerical aperture oil immersion objective lens to reach 69$\%$ collection efficiently in air with the added benefit of the ability to perform Fourier imaging.\cite{Moal2013} Many other approaches have been proposed from which one could single out a few using lenses \cite{Keizer2009,Kuhnke2010,Hoffmann2002,Chen2013}, mirrors, \cite{Freund2011,Khang1999,SUZUKI1999} or optical fibers \cite{Watkins2007,Sakurai2005}. Some of these approaches have light collection efficiencies below 10$\%$, some of them below 1$\%$, limiting very much their practical use. Some systems with higher efficiencies operate only in air or are restricted to specially designed STMs. However, despite all efforts, light collection systems compatible with adapted versions of commercially available LT UHV STM and with collection efficiencies above 50 $\%$ are still sparse.

In this paper, we present the design criteria, some conception aspects, and representative results from an alternative approach to light collection and injection in an LT-UHV STM reaching a collection efficiency of about 72$\%$ of the hemisphere. In our approach, an off-axis parabolic mirror is placed between the sample surface and the STM tip holder. A hole in the mirror allows the tip to reach the sample surface. The tunneling junction is at the focal point of the parabolic mirror and hence the collected light is collimated and sent to the exterior of the UHV chamber through a vacuum viewport. The whole light collection, transmission, and detection scheme is designed to have a high collection efficiency associated with a high spectroscopic resolution with minimum transmission loss. Also, the device is compatible with most commercial Pan STMs after need adaptations. In this work, we show results obtained with the specially modified STM demonstrating its performance at LT and room temperature (RT) on Highly Oriented Pyrolytic Graphite (HOPG), Si(111)-7x7, and WSe$_2$ surfaces. Also, we have performed luminescence measurements in different conditions and samples. First, we show STM-LE under UHV at LT temperature on a gold (111) surface to observe the plasmonic emission from the tunneling junction. We also show \emph{in situ} PL measurements on WSe$_2$ monolayers and spherical core-shell quantum dots as well as \emph{in situ} Raman measurements on HOPG substrate under UHV conditions. In an exfoliated WSe$_2$ monolayers on a gold thin film substrate, we observed excitonic emission and also plasmonic emission from the substrate, both excited by the tunnel current. At RT and 100 K, we observed PL ad STM-CL from defects on a monolayer of h-BN. Finally, with the STM in air, we show STM-LE in exfoliated WSe$_2$ monolayers on a gold substrate. 
 
\section{\label{sec2} Device design for luminescence in STM}

\subsection{\label{sec2.1} General design guidelines}

Among the several possibilities of light collection devices for STMs (mirror, lens, fibers) and also the different STM strategies (Pan design, Beatle design, or others) and operating conditions (LT, RT, UHV, or in Air), some choices and compromises are always made. Detailed discussion about the design of STMs can be found in references \cite{Jchen,Bert} while a review discussing light collection strategies is given in ref \cite{ROSSEL2010}. In this paper, we discuss a solution consisting of the development of a high-performance add-on accessory to suitable Pan STMs. In this sense, for the combined STM and optical device for light collection and injection, we considered the following compromises:

\begin{itemize}
\item The device must be compatible with an STM able of being operated at LT (chilled tip and sample), UHV, or in air; STM tips should be readily exchangeable in vacuum; STM modifications to accommodate the device should be minimal;
\item The device should have a high solid angle, at least higher than 50$\%$ of the hemisphere (collection solid angle > 3 sr), have efficient transmission up to the detector, and display to high spectral resolution (suitable for \emph{in situ} Raman spectroscopy);
\item The optical device for light injection and collection must be compatible with several optical experiments, including PL and Raman spectroscopy, for instance;
\item Any effects of the light collector on the performance of the STM in regular imaging and spectroscopic measurements should be negligible when not in use;
\item The alignment of the light collection system should be sufficiently accurate to assure its performance.
\end{itemize}

A suitable STM design for such light collection and injection device is the Pan design: the tip has coarse approach motion and all scanning axis while the sample has coarse motion in the plane.\cite{Bert} It is very rigid, allows easy tip exchange, and is compatible with low-temperature operation. Pan design also allows sufficient clearance for mirrors with high NA. Using a Pan STM, the mirror is aligned with the tunneling junction as defined by the tip position. Moving the sample does not affect the mirror alignment and positioning regions of interest under the tip is possible without affecting the mirror alignment. In this paper, we discuss the performance of this optical device as installed in a adaped UHV PanScan Flow Cryo STM from RHK Technology. From the optical device point of view, the PanScan has the added advantage that only minor modifications were needed to make it compatible with the insertion of the device (see Sec. ~\ref{sec2.5}).

\begin{figure}
 \centering
\includegraphics{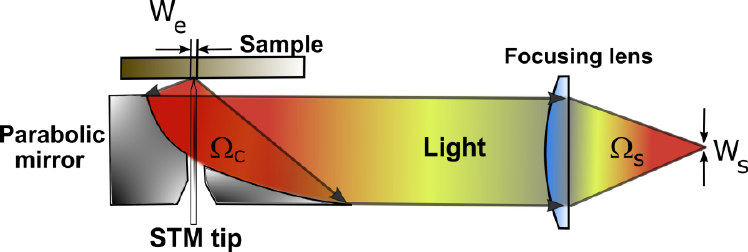} 
\caption{\label{Brightness} Illustration of the basic elements (solid angles and widths) to be considered in the Luminance/\emph{Étendue} conservation.}
\end{figure}

\subsection{\label{sec2.3} Optical design considerations}

Given the definition of the major general design points regarding the STM and the parabolic mirror, the whole light transmission chain must be evaluated. For that, the starting point is the Luminance (also called Radiance or Brightness) conservation theorem together with the \emph{Étendue} conservation theorem.\cite{Bass_OpticsBook_V2,Bass_OpticsBook_V3} The Luminance is related to the amount of power emitted by a source and its spacial and angular distributions and can be expressed as:

\begin{equation}
L=\frac{d\Phi}{cos (\theta) d\Omega_{c} dA} = n^2 \frac{d\Phi}{d \emph{Étendue}_c}
 \label{Eq_Definition}
\end{equation}

Where $L$ is the emitted Luminance collected from the source, $n$ is the refraction index, $\Phi$ is the emitted power, $d\Omega_{c}$ is the differential of collection solid angle, $dA$ is the differential emission area and $\theta$ is the angle between the normal to the differential area and the centroid of the differential solid angle.\cite{Bass_OpticsBook_V2} Since we will consider only the power that is actually collected by the Parabolic mirror, the solid angle considered is already that of the collection mirror. Hence, eq. ~(\ref{Eq_Definition}) refers already to the collected Luminance. The relation of the Luminance (collected from a source) with the \emph{Étendue} of the collection optics is also given in (\ref{Eq_Definition}).\cite{Bass_OpticsBook_V3} The \emph{Étendue} is related to the light gathering and light transferring power of an optical system and/or each one of its elements, as discussed in ref. \cite{Bass_OpticsBook_V1,Bass_OpticsBook_V3}, and describes the trade-offs in terms of angular and spatial distributions when collecting or transferring light. The \emph{Étendue} is defined as

\begin{equation}
\emph{Étendue} = n^2 \int \int cos(\theta) dA d\Omega
 \label{Etendue_def}
\end{equation}
 
and can be seen as the weighted solid angle area product, both for collection and transmission.\cite{Bass_OpticsBook_V3} As this quantity is conserved, it shows that when reducing the solid angle, an increase in area is expected. For clipped Lambertian sources, which is the case that we will consider here as a simple representation of the light emitted by the region being scanned by the STM, the \emph{Étendue} can be expressed as:

\begin{equation}
\emph{Étendue}_{Clip. Lamb.} = n^2 \pi W^2 sin^2\theta^{max}
 \label{Etendue_Lambertian}
\end{equation}

Where $\theta^{max}$ is the maximum angle between the normal of the area $W^2$ and outer points of the optical element collecting or transferring light, and $W$ is the total width of the area emitting or transmitting light (this considers that the system has rotation symmetry which is not the case for the parabolic mirror).\cite{Bass_OpticsBook_V3} From the emission point of view, $W_e$ is the convolution of the used imaging Field of View (region of the sample been scanned by the STM tip), and the carrier spread on the sample (similarly to the beam spread in TEMS or "ionization pears" in SEMs). Therefore, $W_e$ represents the total width from which emitted light should be collected. Note that $\theta^{max}$ is related to the solid angles $\Omega$ under consideration, as indicated in Fig. \ref{Brightness} and its differentials $d\Omega$ in eqs. (\ref{Eq_Definition}) and (\ref{Etendue_def}).

The notion of \emph{Étendue} and Luminance (or Brightness, as typically called in the Electron Microscopy community) and its conservation is important in the case of STM-luminescence experiments (and for cathodoluminescence in electron microscopes). Indeed, as already mentioned, the total emitted power $\Phi$ is typically small. However, this weakness is counterbalanced by the very small emitting area $W^{2}_{e}$ in the definition of Luminance. In other words, an STM excited luminescent source is a bright, but not an intense, emitter. Preserving Luminance (both in terms of power and in terms of coherence) is therefore essential here. At the same time, large collection angles, represented by $\Omega_c$ in Fig. \ref{Brightness} and by $\theta^{max}$ in eq. ~(\ref{Etendue_Lambertian}), are needed to increase the collected power, $\Phi$.

Luminance is conserved in an ideal optical system, but in reality, it can be reduced, for instance, by aberrations, geometrical and chromatic, that induce an increase of spatial and angular distributions. Also, some Luminance is lost in surface reflections and by apertures that reduce $\Phi$, for instance. Therefore, to preserve the collected power (linked to the measured signal intensity) and the Luminance (linked to spectral resolution, see more on this in the following), each optical element up to the spectrometer should be designed carefully. 

Given the Luminance of the source and its conservation, one needs to make sure that the \emph{Étendue} (also called Geometrical Extent) of the optical system is preserved so that none of the collected power is lost. To optimize the spectroscopic optical acquisition chain, it is best to isolate angular and spatial distributions in the spectrometer and collection ends using the \emph{Étendue} for a Lambertian source:

\begin{equation}
\frac{\Phi}{\pi L}= W^2_e sin^2\theta^{max}_c \leq W^2_s sin^2\theta^{max}_s =\frac{\emph{Étendue}_s}{\pi n^2}
 \label{OpticalSystem}
\end{equation}

Where $\theta^{max}_{s}$ is the angular acceptance at the spectrometer entrance and $W_s$ is the light beam width in the spectrometer dispersive direction. The inequality $\leq$ means that the product $W^2_s sin^2\theta^{max}_s$ should be equal or larger than $W^2_e sin^2\theta^{max}_c$ to accept all collected light. Simply put, eq. ~(\ref{OpticalSystem}) means that there is a trade-off between collection efficiency, related to $\theta^{max}_{c}$, and field of view, given by $W_{e}$, on the sample side, with respect to the spectrometer acceptance angle, $\theta^{max}_{s}$, and slit width, $W_{s}$, on the spectrometer side. Also, eq. (\ref{OpticalSystem}) means therefore that the spectrometer \emph{Étendue} is a limiting factor to the compromise between light collection efficiency and field of view. Fig.~\ref{Brightness} illustrates the solid angles and widths in eq. (\ref{OpticalSystem}). Note that the angles in eq. (\ref{OpticalSystem}) refer to conical angular ranges in 3 dimensions and, therefore, relate to the solid angles in space.

\begin{figure*}
\includegraphics[width=5.4 true in] 
{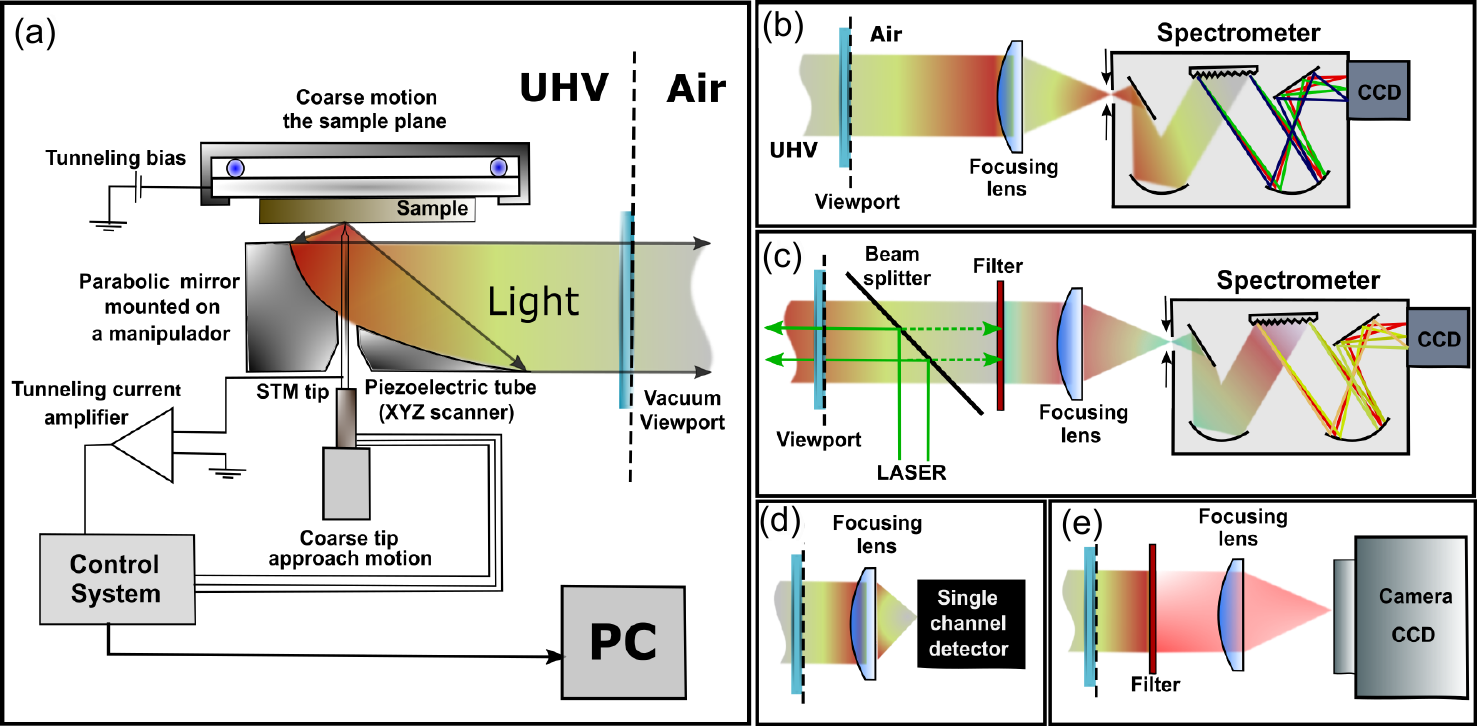}
\caption{\label{OpticalExperiments} (a) Scheme of the parabolic mirror inside the STM directing collected rays perpendicularly to the direction of the tip. (b-e) Possible experiments: in (b) all light is directed to the entrance of a spectrometer; in (c) a beam splitter is inserted in the parallel beam path and \email{in situ} PL/Raman can be performed; in (d) the light is injected in a large area single channel detector; and in (e) the angular distribution of the emission is recorded by a CCD positioned directly in front of the parallel light beam.
}
\end{figure*}

In practical terms, most light spectrometers use parallel detectors such as CCDs, and their pixel size together with the spectrometer point spread function (PSF) and magnification define the relation between the entrance slit width and the spectrometer resolution (for a given grating). Indeed, the spectral resolution depends on several parameters, including some intrinsic to the spectrometer (NA, aberrations, grating), some extrinsic (especially the size of the beam at the entrance aperture). The latter needs to be optimized for maximum spectral resolution. It is, therefore, a matter of choosing a target resolution (width of peaks on the CCD) and then injecting the light already with the desired light beam width, $W_{s}$. Note that eq. ~(\ref{OpticalSystem}) shows the link between the Luminance of the source (left-hand side) and the spectrometer entrance slit width, that is its resolution (right-hand side). Typically, for $25\hspace{2mm} \mu$m wide pixels, a light beam width of about 100 $\mu$m entering the spectrometer is already sufficiently narrow to provide a spectral resolution near the maximum attainable (without even using the slits), since a typical spectrometer magnification is near 1 and its point spread function has a full width at half maximum (FWHM) of about two pixels.

In our setup, for a 600 gr/mm grating, the spectrometer resolution is about 0.22 nm for 10 $\mu$m entrance slits and 0.5 nm entering with a beam width of 100 $\mu$m (and entrance slits fully open). Such spectral resolution is sufficient for most experiments with semiconductors and can be further increased by changing the grating. Closing the slits to reduce $W_{s}$ is still possible but it would reduce the spectrometer \emph{Étendue}, causing light loss. Considering a spectrometer with NA=0.12, that is with a maximum acceptance angle of 7\degree, and using $W_{s} = 100$ $\mu$m, and considering a parabolic mirror with a high numeric aperture with $\theta^{max}_c$ about 80° (see Sec.~\ref{sec2.4}), one has:

\begin{equation}
W_e \leq \frac{ W_s sin\theta_s}{sin\theta^{max}_c} \cong 12 \text{ }\mu\text{m} 
 \label{eq4}
\end{equation}

This means that, provided all light emission comes from a region smaller than $W^2_{e} = 12^2$ $\mu$m$^2$, a high proportion of the emitted light (>70$\%$, see Sec.~\ref{sec2.4}) can be collected and transmitted to the spectrometer without loss into a width $W_{s} = 100$ $\mu$m. Note that if the emission width is larger (or the mirror is misaligned, what is equivalent), the Luminance would be reduced accordingly and signal would be lost. Therefore, the mirror alignment must be accurate in the sub-micrometer range and only a region of a few micrometers can be explored on the sample plane. One must note that in some situations, like in some plasmonic applications, keeping a high spectral resolution is not mandatory. In this case, a larger width at the spectrometer entrance could be used which would ultimately give more flexibility on the light collection optics. To reach larger regions, the mirror could be scanned together with the tip as demonstrated in ref. \cite{Edelmann2018} or the sample could be scanned.\cite{Edwards2011} 

It is important to note that this calculation does not depend on the specific microscope holding the light collection optics and therefore it applies to SEMs and TEMs performing cathodoluminescence and STMs performing STM-LE or other. Another relation between collection efficiency, field of view, entrance slits width, and spectrometer numerical aperture was shown in ref. \cite{Edwards2011} which was derived using a different approach in two dimensions and provided different values but a very similar trend. Finally, these calculations must be considered as an upper limit for an ideal system with rotation symmetry. A given particular optical system may not reach this value and may have a narrower field of view or alignment tolerance. In particular, off-axis parabolic mirrors are known for their low tolerance to misalignment, and simulations (see in sec. \ref{sec2.4}) indicate that $W_e$ in our system is smaller than predicted in this model. 

A general scheme of the STM together with the light collector is presented in Fig.~\ref{OpticalExperiments}(a). Basically, in the Pan design, the coarse motion of the tip base moves the tip upwards until a tunneling current is observed. At this point, the electronics of the STM control the tip-sample distance using the scanner tube to maintain a constant current. A lateral scan can be performed for imaging in constant tunneling current mode in which the current is held constant by adjusting the tip-sample distance along the scan. Luminescence induced by the tunneling current illuminates the parabolic mirror which directs the light beam as parallel rays towards the outside of the thermal shields and also outside of the UHV chamber to a dedicated collection optics. Note that this approach allows one to exploit all advantages of free space optics for different experiments and for light injection as well. Finally, the coupling to the spectrometer entrance can be accomplished by free optics using only mirrors and lens or additionally using optical fibers.\cite{Chen2013,KociackPat}

\subsection{\label{sec2.2} Light collection/injection possibilities}

Considering that the parabolic mirror reflects all gathered light as a parallel beam to the exterior of the UHV chamber, it is a matter of working with this beam in free optics in air. Therefore, outside the vacuum viewport and on top of an optical table, several experiments are possible. Typically, tunneling current induced luminescence (STM-LE) is sent to the entrance of an optical spectrometer with a CCD detector, as shown in Fig.~\ref{OpticalExperiments}(b). In this way, STM-LE spectra can be acquired for a given tip position and tunneling condition.\old{In this case, the acquired luminescence spectra originates from a small sample region} Moreover, spectral images can be acquired by synchronizing the STM tip scanning with the spectra acquisition. When working under UHV conditions, it is also possible to operate the STM in Field Emission (FE) mode to perform cathodoluminescence experiments (STM-CL). In this case, the STM tip is retracted 100-200 nm from the sample surface and an external voltage source (0-500) V is used to produce an electron beam and excite the sample by electron bombardment. \cite{JuanJose,Watanabe2013CL,RudolphM2014,Stavale2013,Stavale2014}

Alternatively, by adding a laser on the optical table, light can be injected to perform \emph{in situ} PL and Raman experiments (with or without the use of the STM tip), as indicated in Fig.~\ref{OpticalExperiments}(c), similarly to ref \cite{UEHARA2014}. Tip effects can be explored by doing tip-enhanced PL or tip-enhanced Raman spectroscopy (TERS) with appropriate tips, as demonstrated and reviewed in \cite{Chen2014,STANCIU2008,Hartschuh2008,Sheng2018}. Interestingly, parabolic mirrors are very good in focusing laser beams and hence very suitable for these techniques.\cite{Drechsler2001,STANCIU2008,Zhang2009,Debus2003} PL can also be useful, for instance, to check that a given sample is optically active \emph{in situ}. The laser spot size is estimated as smaller than 2 $\mu m$ as observed from sample damage (dark spots observed on samples after PL experiments). 

Another possibility, shown in Fig.~\ref{OpticalExperiments}(d), consists in directing the light to the entrance of a large area single channel detector, like a photo-multiplier tube (PMT) or a photodiode, which can be useful to perform luminescence mapping.\cite{Pommier2019} This mode is also interesting for QE measurements. Indeed, some single channel detectors have well documented QE and it is possible to measure the quantum yield of a given sample in a specific tunneling condition. This is relevant when absolute quantum yield is required and a good mean of comparing data among different setups. Also, in this setup, only the viewport and one lens are placed between and mirror and a large area detector (which minimizes any possible misalignment loss). This ensures low and controlled light loss such that detector counts can be accurately related to emitted photons. This is not exactly the case for the setup in Fig. ~\ref{OpticalExperiments}(b) since some elements in the optical path are less accurately known and, most importantly, the CCD QE is not always well-known.~\cite{Sperlich_2013}

Finally, by placing a CCD camera in front of the mirror, as shown in Fig.~\ref{OpticalExperiments}(e), angle-resolved emission patterns can be recorded, similarly as has been done by Le Moal, et al. \cite{Moal2013} and indicated by Romero, et al.\cite{Romero2006}. This strategy has been extensively used in SEMs in A. Polman Group.~\cite{Osorio2016,Coenen2011}. This mode is particularly relevant to assist the understanding of the light emission mechanism.~\cite{Pommier2019} As shown in Fig.~\ref{OpticalExperiments}(e), a filter (and also a polarizer) can be used to distinguish among different light signals. 

Another possible experiment is coincidence detection: using a beam splitter and two single channel detectors\cite{Zhang2017,Leon2019,Merino2015}. It must be noted that such experiments are possible with off-axis parabolic mirrors as demonstrated in STEM and SEM \cite{KOCIAK2017,CoenenCentury,Meuret2018} and have also been demonstrated in STM-LE (using lens or mirrors)~\cite{CoenenCentury}.

\subsection{\label{sec2.4} Optical Simulations}

\begin{figure}
 \centering
\includegraphics[width=3.0true in]{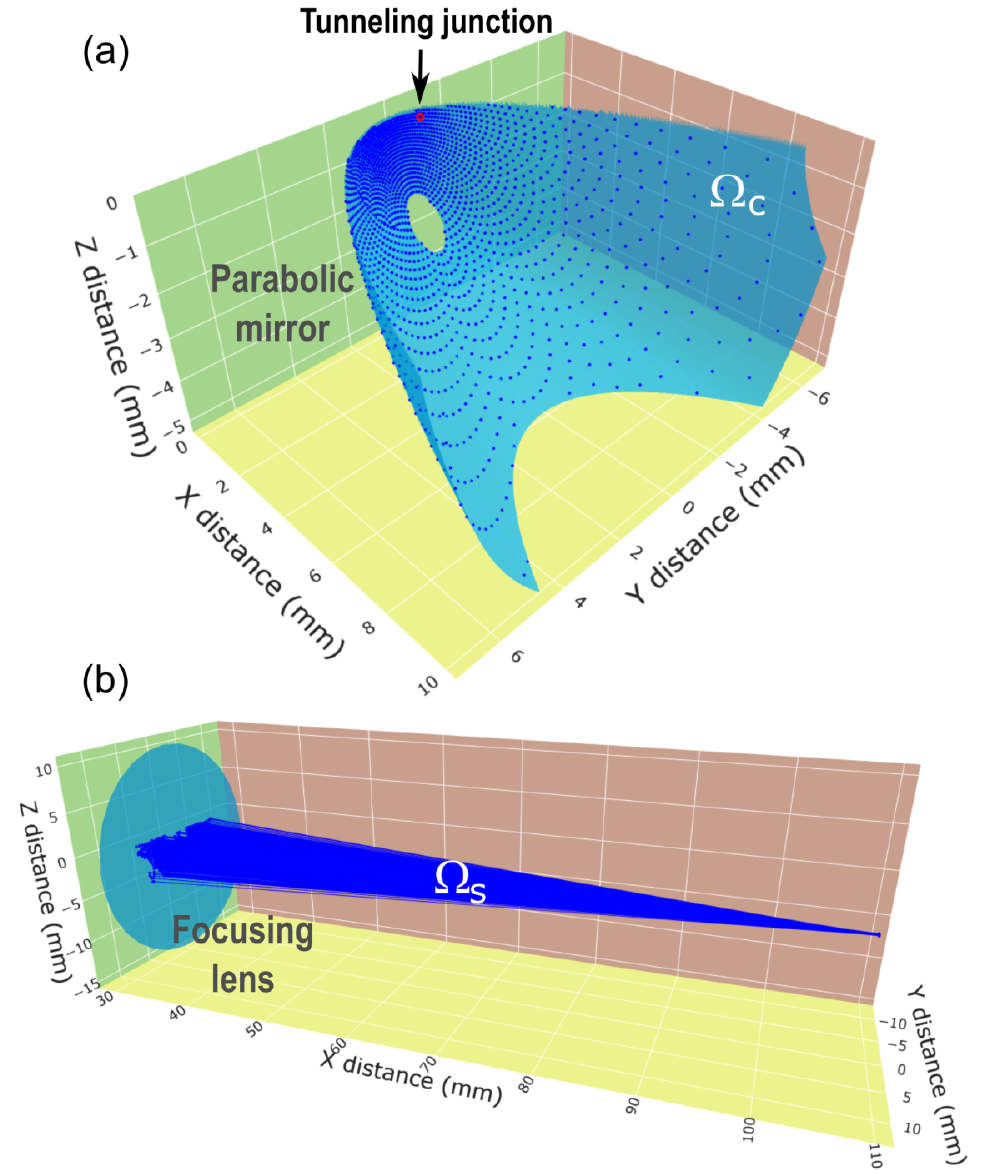}
\caption{\label{RayTracing} { Typical visualization to inspect the transmission of isotropic light rays emitted by the tunneling junction (red point in (a)) up to the spectrometer entrance. (a) Emission from the isotropic source is collected by the parabolic mirror. Each blue dot indicates where a given ray from the source hit the mirror (semitransparent blue surface). (b) Rays are focused by a lens (semitransparent blue surface) into the spectrometer entrance. Each line indicates the path light ray that was collected by the mirror and send to the lens and is shown to focus into a small region corresponding to the position of the spectrometer entrance.}}
\end{figure}

Taking into account all design considerations that defined the key aspects of the light detection device, it is important to determine its efficiency and usability by simulations. Indeed, the model presented in eq. (\ref{OpticalSystem}) refers to systems with rotation symmetry, considers a clipped Lambertian source, and considers that the optics is in fact aligned. Simulations can evaluate the system for an isotropic source, can consider the actual mirror shape, and consider the misalignment properly, that is: the source is outside of the focal point (and not an extended source). For that, we have performed optical ray trace simulations using homemade software.\cite{RayTraceBook} Here, simulations consider the light source as a dimensionless isotropic emitter. This source model represents an individual emitter for a fixed STM tip position. The source is moved on the surface plane to account for the scanning and charge carrier spread. Diffraction effects are not taken into account.

Optical ray trace simulations follow individual light rays from the isotropic emitter along the optical path (parabolic mirror and focusing lens) up to the spectrometer entrance (or the optical fiber entrance). The code considers a large number of rays for statistical relevance and provides detailed information on the outcome of each light ray emitted from the source. From the simulations, the collection solid angle of the mirror employed was determined to be 72$\%$ of $2\pi$, that is, 72$\%$ of the hemisphere or $4.5$ sr, with an $\theta_{max}$ of about 80\degree. This compares well to recent publications in the field such as 75$\%$ in ref. \cite{Edelmann2018} for a similar LT UHV STM or 69$\%$ in ref. \cite{Moal2013} for an air STM also capable of Fourier imaging. Moreover, as discussed previously, the initial \emph{Étendue} is preserved up to the spectrometer in the proposed design, which gives high spectral resolution (0.5 nm as discussed in Sec. ~\ref{sec2.2}) and high transmission efficiency of ($\approx$ 50$\%$ as discussed in Sec. \ref{sec2.5}).

Fig.~\ref{RayTracing} shows one example of a visualization of such simulations. Fig.~\ref{RayTracing}(a) shows the position where light rays from an isotropic source hit the mirror surface. After hitting the mirror, the light follows as parallel rays to the focusing lens. In Fig.~\ref{RayTracing}(b), the lens focuses the rays in the optical fibers (equivalent to the spectrometer entrance). This kind of visualization can be used to inspect the simulation and understand the light path from the source, for instance.

\begin{figure}
 \centering
\includegraphics[width=2.6 true in]{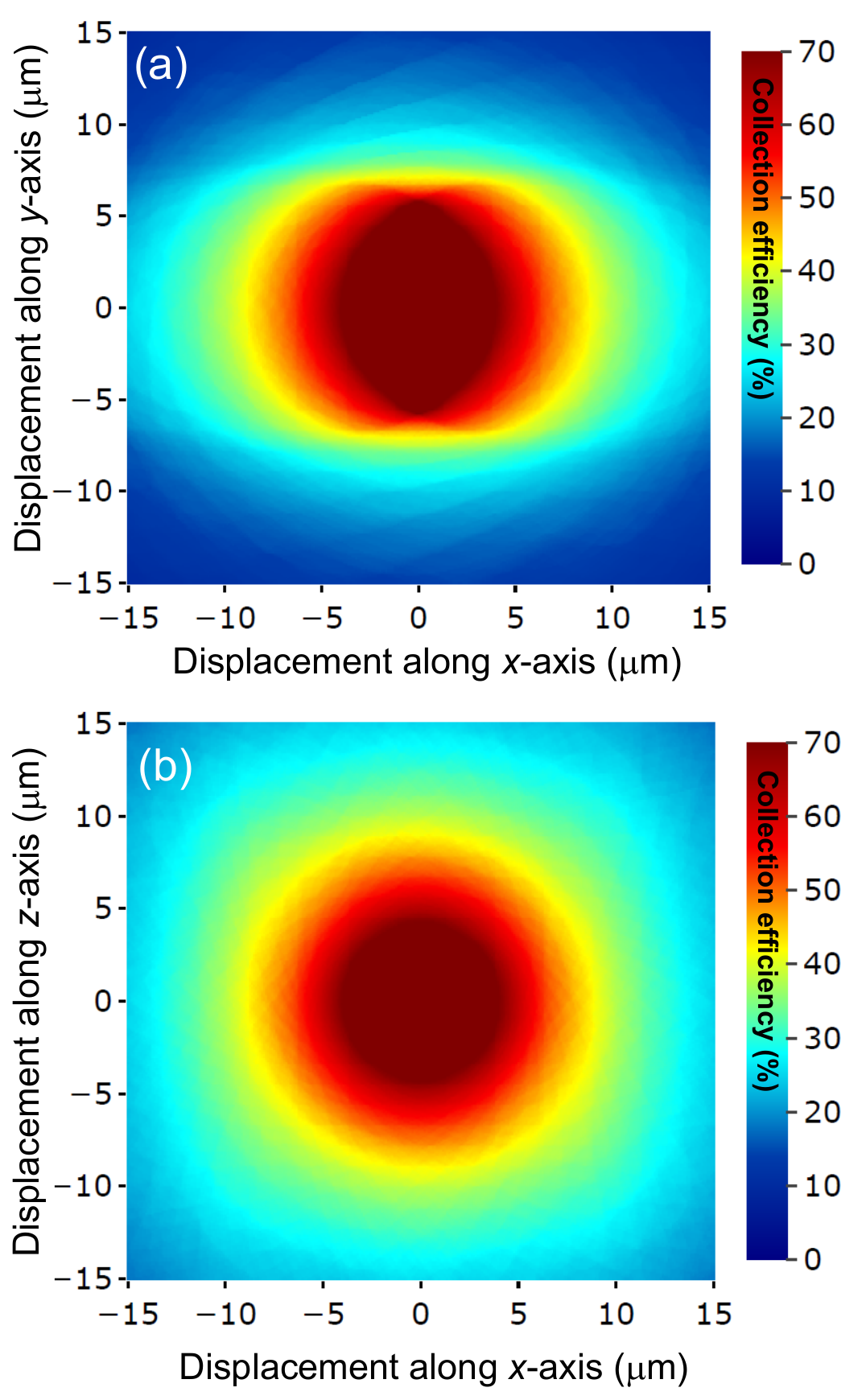}
\caption{\label{HeatMap} Heat map that shows the collection efficiency (in $\%$) as a function of the position of a point emitter in the sample plane, x-y, (a) and out of the plane, z-x, (b). The dark red region has maximum collection and corresponds to the W$_e$ region. }
\end{figure}

The result of another relevant simulation is the region on the sample that can be scanned without loss of collection efficiency. Indeed, as observed in eq. (\ref{eq4}), the higher the numerical aperture (collection solid angle), the smaller the field of view for a given \emph{Étendue} (and similarly for the depth of focus).\cite{Edwards2011} As mentioned, given the mirror geometry, calculating the region of emission with maximum \emph{Étendue} also provides information about the misalignment the mirror can sustain without significant collection loss as moving the tip is equivalent to moving the mirror. A typical result of such simulation in the case of our system is shown in Fig.~\ref{HeatMap}. 

Figures~\ref{HeatMap}(a) and ~\ref{HeatMap}(b) show the area which preserves optimum collection (dark red area) on our system. They indicate that the mirror and the light source can be about 5 $\mu m$ apart in any direction without loss of efficiency. This indicates both the region that can be scanned by the STM tip and the mirror alignment tolerance. Eq. (\ref{eq4}) can not predict this region accurately as it does not take into account some aspects of the actual optical system. The most relevant difference is that eq. (\ref{eq4}) considers an extended source with area $W_e^2$ (Lambertian source) while simulation in Fig.~\ref{HeatMap} considers a point source (isotropic emitter) that is moved in the same plane (sample surface plane), which better represents the actual physical system (particularly for charge carrier diffusion length smaller than 1 $\mu m$). Another difference between the model and optical system is the absence of rotational symmetry (needed in the model). Also, the simulations consider the exact mirror shape while the simulations use only its angular aperture $\theta_{max}$.

The results in Fig.~\ref{HeatMap} also recall that, again similarly with STEM-CL, STM-LE is a method that excites very locally but collects globally.\cite{KOCIAK2017,Zagonel2012,Zagonel2016} That means that one knows where carriers where injected or where the excitation source is with great precision (possibly in the atomic scale), however the emission can happen more than a micrometer away as already observed in TMDs.\cite{Pommier2019} In our case, the whole area in Fig.~\ref{HeatMap} is collected and hence the emission could come from anywhere in that area. This is very different from confocal optical microscopy, for instance, in which both injection and collection are related to a small area.

\subsection{\label{sec2.5}{Optical device implementation}}

\begin{figure}
 \centering
\includegraphics[width=2.0true in]{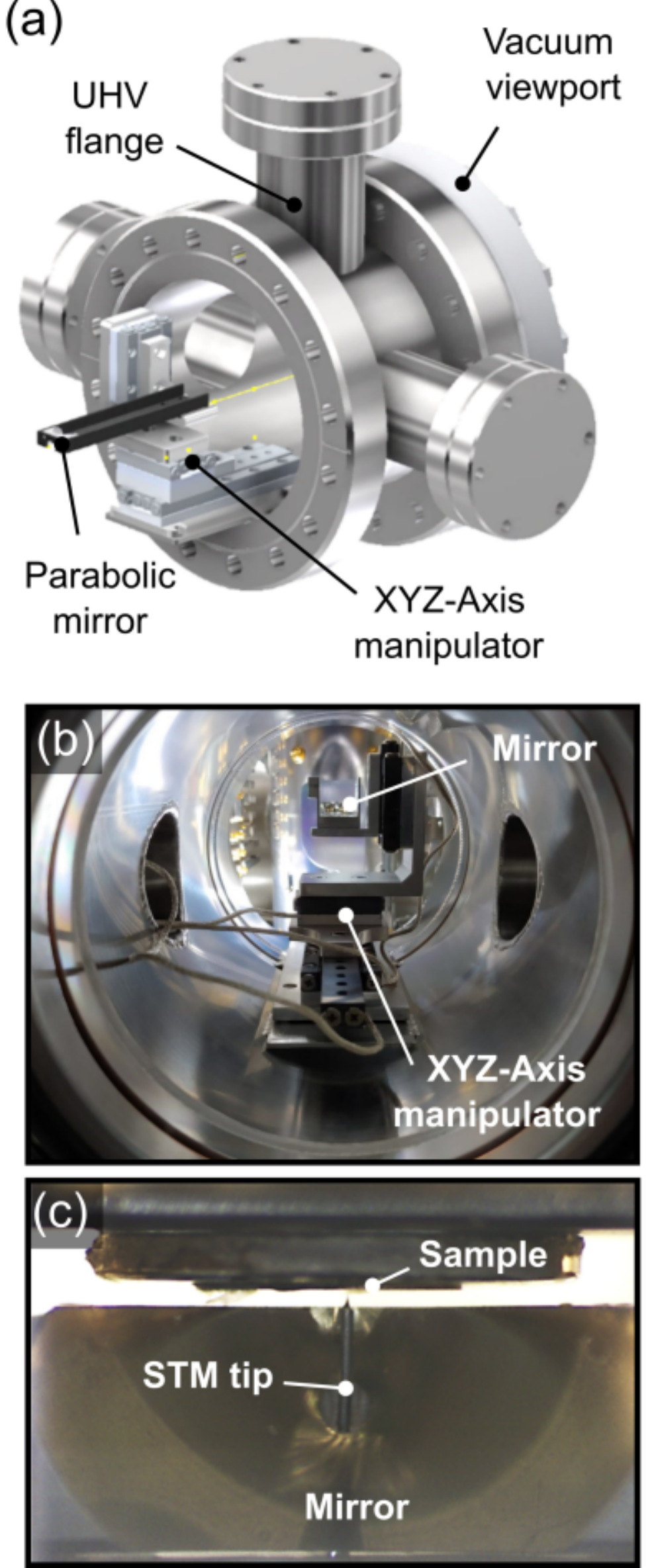} 
\caption{\label{OpticalDevice} (a) Optical device chamber with the manipulator and the mirror in a standby position. (b) A picture of the manipulator as observed from the rear. (c) A picture of the mirror showing an STM tip and the sample. }
\end{figure}

Considering the design criteria in Sec.~\ref{sec2.1} and the alignment demands in Sec.~\ref{sec2.3} and ~\ref{sec2.4}, we chose to use a piezo-driven manipulator with a long arm that holds the mirror. To implement it, we build a small chamber, as shown in Fig.~\ref{OpticalDevice}(a). Such manipulator can place the tunneling junction at the focal point of the mirror and also retract the mirror outside the STM thermal shields. When the mirror is retracted, the STM thermal shields can be fully closed as usual. In this design, the manipulator and its mirror never enter in mechanical contact with the STM body or thermal shields. Fig.~\ref{OpticalDevice}(a) displays a  computer-aided design (CAD) drawing of the chamber where the manipulator is installed. Fig.~\ref{OpticalDevice}(b) shows a picture of the manipulator from the rear of the chamber (viewport side). In Fig.~\ref{OpticalDevice}(c), it is possible to see a picture of the mirror showing the STM tip passing through the mirror hole and the sample on the top. It is worth mentioning that the tip length has to be a bit longer than usual to pass through the mirror. Therefore, we adapted the tip holder and we could not notice a significant difference in image quality using the standard tip holder with short tips or the adapted tip holder with slightly longer tips (see Sec. \ref{sec3}).

\begin{figure}
 \centering
\includegraphics[width=2.8true in]
{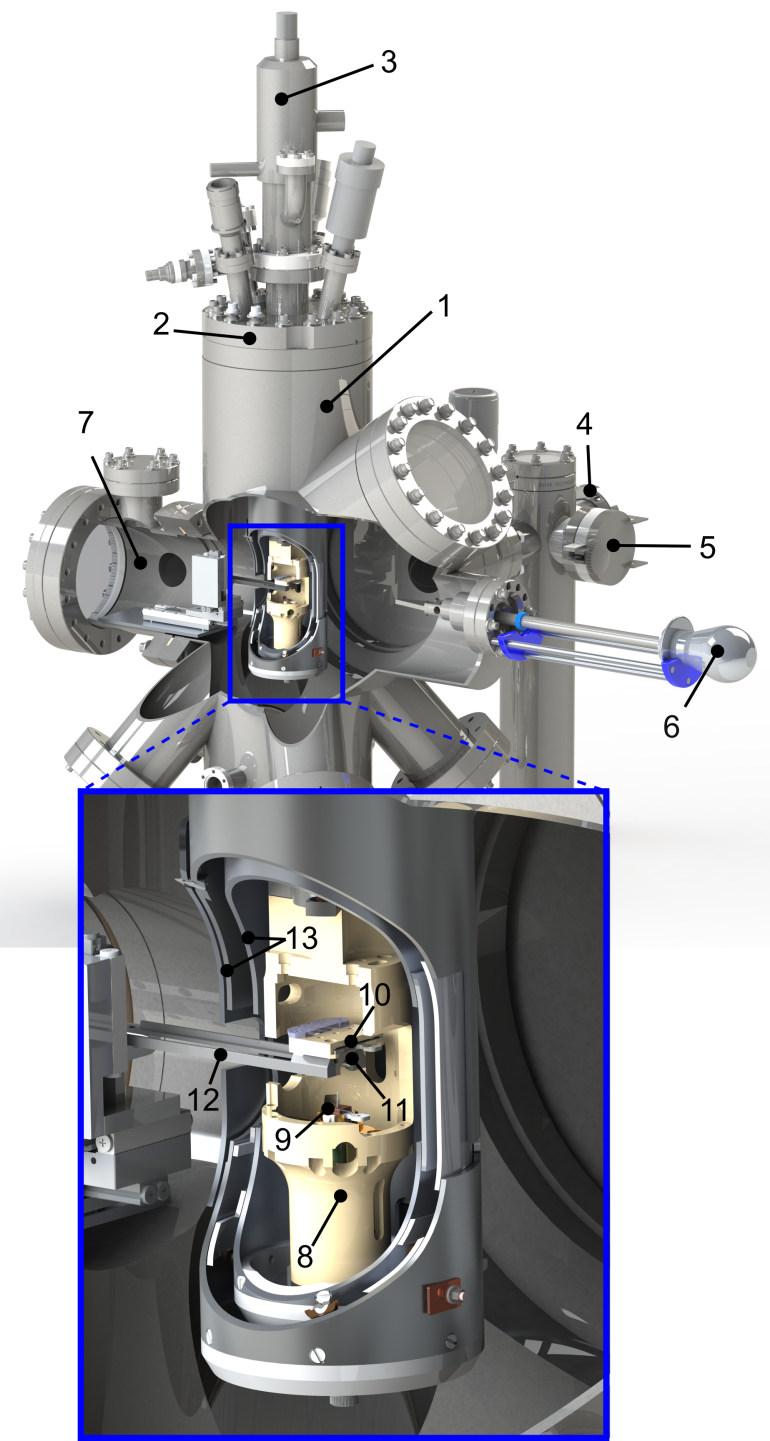}
\caption{\label{STMChamber} CAD view of the RHK STM body and thermal shield revealing the opening and showing the manipulator of the optical device holding the mirror in a measurement position. (1) UHV main chamber, (2) STM flange, (3) cryostat, (4) transfer arm, (5) load-lock chamber, (6) wobble stick, (7) optical device chamber, (8) STM head, (9) STM tip, (10) sample, (11) mirror, (12) manipulator arm (mirror support), and (13) thermal shields.}
\end{figure}

Modifications were needed to adapt the RHK PanScan STM body, thermal shields, and chamber to receive the optical device. Openings on the rear of the STM and thermal shields were added or adapted to allow the positioning of the mirror. Fig.~\ref{STMChamber} shows the optical device chamber attached to the STM main chamber and, in the insert, the mirror in position with the arm of the manipulator passing through the thermal shield and STM body. 

The fact that the mirror is mechanically decoupled from the STM has advantages and drawbacks. When not in use, surely, the performance of the STM is not affected by the optical device. However, inserting the mirror in position will drive thermal instabilities. That is basically because the STM has a weak thermal link to the cryostat to avoid strong mechanical coupling. Hence, at the moment, the operation is more stable when the optical device is used with the STM at room temperature. In some physical systems such as WSe$_2$ monolayers, that is, in fact, the best temperature to operate.~\cite{Samadi2018, Withers2015} Nevertheless, using the optical system at LT is feasible depending on the tolerance to thermal drifts one can bear. Also, simple modifications can be done to cool the mirror and allow stable measurements at low-temperature as well.

\begin{figure}
 \centering
\includegraphics[width=2.4true in]
{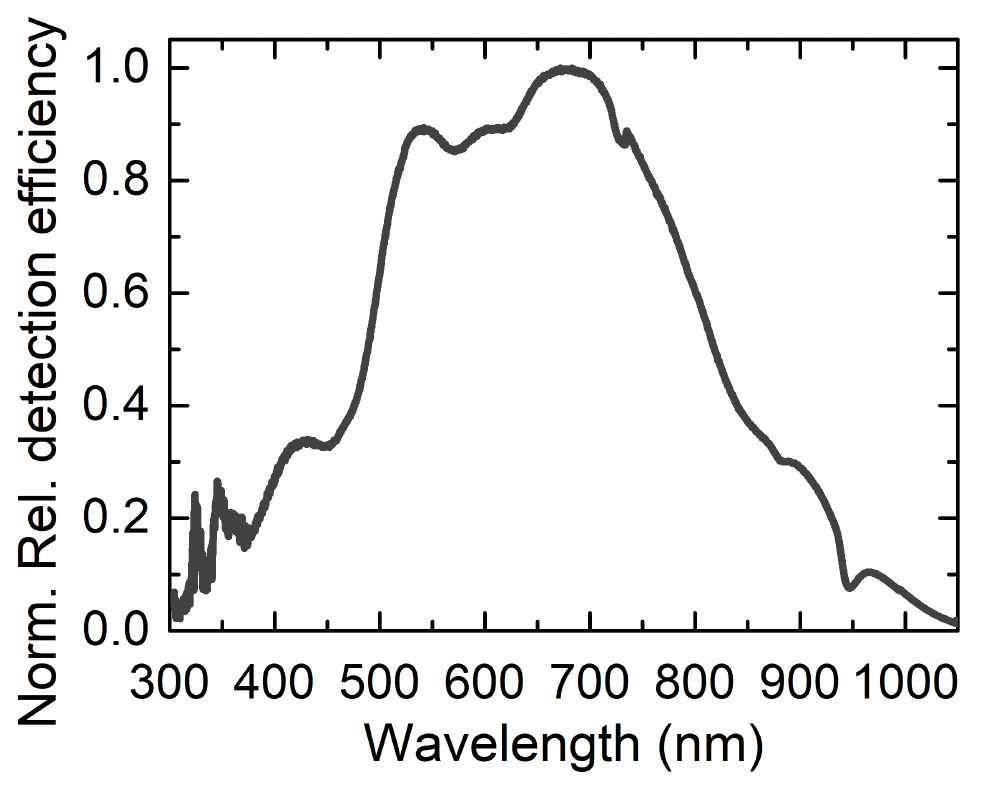}
\caption{\label{InstrumentResponseFunction} Instrument response function for a 300 gr/mm diffraction grating blazed at 600 nm.}
\end{figure}

Outside the viewport indicated in the drawing of Fig.~\ref{OpticalDevice}(a), several optical experiments can be realized, as discussed in sec. ~\ref{sec2.3} and shown in Fig.~\ref{OpticalExperiments}. Most importantly, for spectroscopy, the light collection system to perform STM-LE contains only 3 refractive optical elements (the vacuum viewport, one lens, and the optical fiber), and hence surface reflection loss is kept small. The total collection/transmission efficiency of the system at 750 nm can be estimated as 50$\%$ (considering from the collection inside the STM up to reaching the spectrometer CCD but without considering the CCD QE) with a spectral resolution of 0.5 nm (for a 600 gr$/$mm grating). In this calculation, we considered initial collection (72$\%$), refractive surface reflections (3$\%$ each), and grating efficiency (80$\%$). Other gratings can be used with higher groove density (for Raman spectroscopy, for instance) or lower groove density (for spectroscopy of plasmonic modes, for instance). Note that the careful matching of all optical elements and their correct alignment, as discussed in ~\ref{sec2.3}, allowed the reduction of light loss to a minimum and reaching such collection/transmission efficiency with 0.5 nm spectral resolution (at 750 nm). The efficiency from the collection (mirror) up to the light beam that will hit the CCD detector with a given resolution is a meaningful way to evaluate the performance of a spectroscopy light gathering system. 

\begin{figure*}
\includegraphics[width=4.8 true in]
{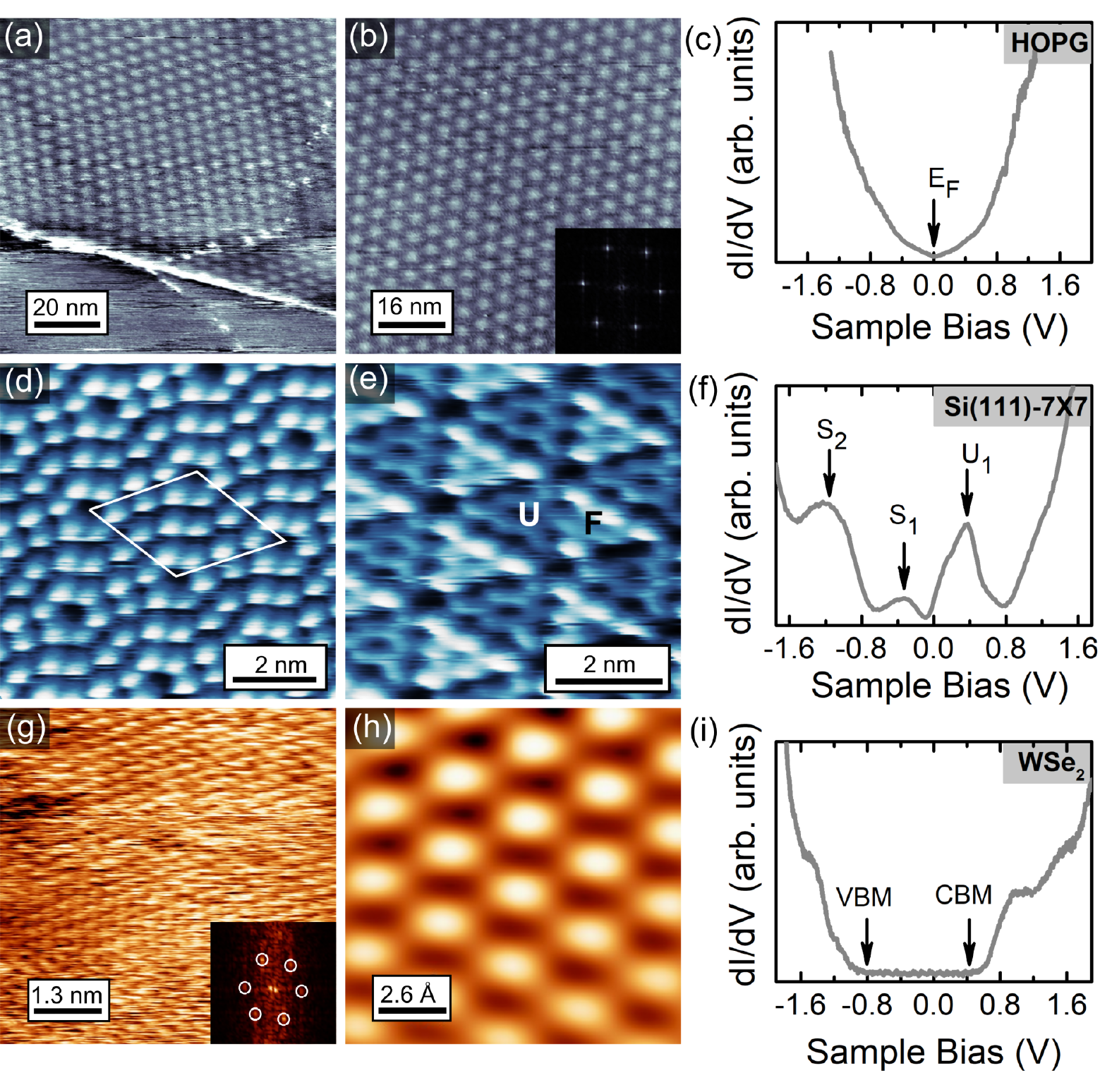}
\caption{\label{STMPerformance} STM images of Moiré pattern in HOPG measured at 12 K and with tunneling parameters of (a) 1.5 V, 300 pA and (b) 1.5 V, 500 pA. Inset: Fast Fourier Transform (FFT) pattern. (c) STS curve measured on HOPG (40 mV, 0.65 kHz). (d) STM image for the empty (0.7 V, 200 pA) and (e) filled (-1.8 V, 200 pA) states of the Si(111)-7x7 surface reconstruction at 80 K. (f) STS curve of Si(111)-7x7 (80 mV, 0.80 kHz). (g) STM image with atomic resolution of the WSe$_2$ surface at 80 K (-3.8 V, -500 pA). Insert: FFT image. (h) FFT filtered image. (i) STS curve obtained in WSe$_2$ (40 mV, 0.65 kHz). All the STS curves were taken at 80 K.}
\end{figure*}

Due to the wavelength dependent efficiency of each optical element, the (relative) spectrometer response as a function of the wavelength has to be determined experimentally. This is particularly important when dealing with large spectra, like those of plasmonic emission, or spectra ranging in spectral regions with significant efficiency variation. Fig.~\ref{InstrumentResponseFunction} shows a typical instrument response function of our setup in its current configuration. The efficiency increase from 300 nm to 600 nm is mostly due to the grating (with some features due to the CCD) while the fall from 700 nm to 1000 nm is mostly due to the CCD. Some absorption regions are noted at $\approx$720 nm, $\approx$ 880 nm, and $\approx$ 940 nm and are due to absorption in the coupling optical fiber. This curve can be used to correct spectra and compensate for the different efficiencies along the available spectral range. STM-LE spectra in Sec. ~\ref{sec4.1} were corrected using such a curve.

\section{\label{sec3} STM performance}

We evaluated the performance of the STM after the modifications mentioned previously by performing STM images and STS on some samples at hand. These tests were performed after the installation of the light collection/injection device and the usual bake-out of the whole system (including the optical device). The chamber pressure remained near 2x10$^{-11}$ mbar after the installation. The results presented in this section were obtained with the mirror fully retracted and the thermal shields fully closed. That is the usual condition for STM/STS operation. The typical temperatures using LHe and LN$_2$ as coolant were 12 K and 80 K, respectively. These temperatures were fairly easy to reach and stable to maintain using a temperature controller. Fig.~\ref{STMPerformance} shows a set of STM images and STS curves obtained operating the STM at LT. The STM images were acquired in constant-current mode with electrochemically etched tungsten tips, and using the adapted tip holder with slightly longer tips (suitable for luminescence experiments). For image processing, the WSXM ~\cite{WSXM2007} and Gwyddion ~\cite{Gwyddion} software were used. The STS curves were taken using a lock-in after obtaining stable topographic images.~\cite{Bert} 

Fig.~\ref{STMPerformance}(a)-(c) show STM/STS results obtained in a HOPG sample and operating the microscope at 12 K. Before the measurements, the HOPG crystal has been cleaved in air using an adhesive tape and promptly introduced into the load-lock chamber. Fig.~\ref{STMPerformance}(a) shows the Moiré pattern usually found in the HOPG surfaces.\cite{Wong2010, PATIL201755, Morell2015, Yin2014,Flipse2009} Moiré patterns are formed by the relative rotation between two graphene layers and its periodicity can be calculated by the equation $D=a/(2sin(\theta/2))$, where $a$=0.246 nm is the lattice parameter for graphene and $\theta$ is the rotation angle between graphene layers. For the structure in Fig.~\ref{STMPerformance}(b), the periodicity $D$ was obtained by calculating the average distance between bright spots using different height profiles and also by superimposing a hexagonal lattice to the STM image. A periodicity of $D= 5.0$ nm has been determined which gives a rotation angle $\theta$ = 2.8\degree. Additionally, Fig.~\ref{STMPerformance}(c) displays an STS curve taken at 80 K. The result is the expected for graphite, this is, a V-shape LDOS symmetric with respect to the Fermi level \cite{Yin2014,BYSZEWSKI200325,Teobaldi2012,Flipse2009}.

The Si(111)-7x7 surface reconstruction was imaged with atomic resolution at 80 K and the results are shown in Fig.~\ref{STMPerformance}(d)-(f). The surface was prepared under UHV conditions by flash annealing of a p-type silicon (111) sample. STM images with atomic resolution for both empty and filled surface states \cite{Bert, Oura2003, Myslive2006} are presented in Fig.~\ref{STMPerformance}(d) and (e), respectively. In agreement with the well-known Dimer-Adatom-Stacking Fault model ~\cite{Takayanagi1985, WangJing2018}, in Fig.~\ref{STMPerformance}(d) the twelve adatoms inside the unit cell are resolved, while Fig.~\ref{STMPerformance}(e) shows the different heights of the faulted (F) and unfaulted (U) part of the unit cell. The LDOS of this sample is given by the STS curve shown in Fig.~\ref{STMPerformance}(f), where three electronic states are clearly resolved: one electronic state (U$_1$) at 0.4 V above the Fermi level and two electronic states at -0.3 V (S$_1$) and -1,2 V (S$_2$) below the Fermi level. U$_1$ and S$_1$ are states localized in the adatoms and the S$_2$ in the rest atoms. ~\cite{Bert, Oura2003, Myslive2006, Odobescu_2012, WangJing2018}

Finally, also at 80 K, we carried out STM/STS measurements on a thick micro flake of WSe$_2$ mechanically exfoliated and transferred on a p-doped silicon substrate. Fig.~\ref{STMPerformance}(g) shows an STM image with atomic resolution of the WSe$_2$ surface. In the inset on Fig.~\ref{STMPerformance}(g), an FFT pattern shows the hexagonal periodicity of the structure. Fig.~\ref{STMPerformance}(h) displays an FFT filtered image revealing the hexagonal atomic pattern clearly and in agreement with already reported images.\cite{Pommier2019, Ponomarev2018} The STS curve in Fig.~\ref{STMPerformance}(i) is typical for a semiconducting sample, where the zero dI/dV region is associated to the electronic band gap of about 1.3 eV, as expected for this material ~\cite{McDonnell2014, Addou2016}, and the Fermi level position near the conduction band minimum indicates light n-type doping.

\section{\label{sec4} Luminescence experiments inside the STM under UHV} 

In the following, luminescence signals (STM-LE, STM-CL, \emph{in situ} PL, and \emph{in situ} Raman) observed from a metallic surface, from semiconducting systems (Core-shell QDots and WSe$_2$), and from HOPG and monolayer h-BN are shown as proof of principle of our implementation of the light collection/injection device discussed in Sec.~\ref{sec2}. The results display the functionality of the light injection and collection device and the potential for forthcoming applications in the study of semiconductor nanostructures. In this section, all the luminescence signals were recorded with the sample inside the RHK Technology PanScan STM and under UHV environment at RT and LT.

\subsection{\label{sec4.1}{Plasmon emission on Gold(111) Surface}}

Metallic surfaces are commonly used to observe the plasmonic light emission in STM-LE experiments. The light emission is originated by the radiative decay of plasmon modes localized in the tunneling junction and excited by inelastic tunneling processes.\cite{Kuhnke2017, Edelmann2018,ROSSEL2010} Plasmon confinement in the cavity formed by a sharp tip and a metallic surface has emerged as an excellent tool for exploring the light-matter interaction at the nanoscale, and has been employed to measure single-molecule Raman spectroscopy, \cite{RZhang2013} for instance. 

\begin{figure}[H]
 \centering
\includegraphics [width=2.453true in]{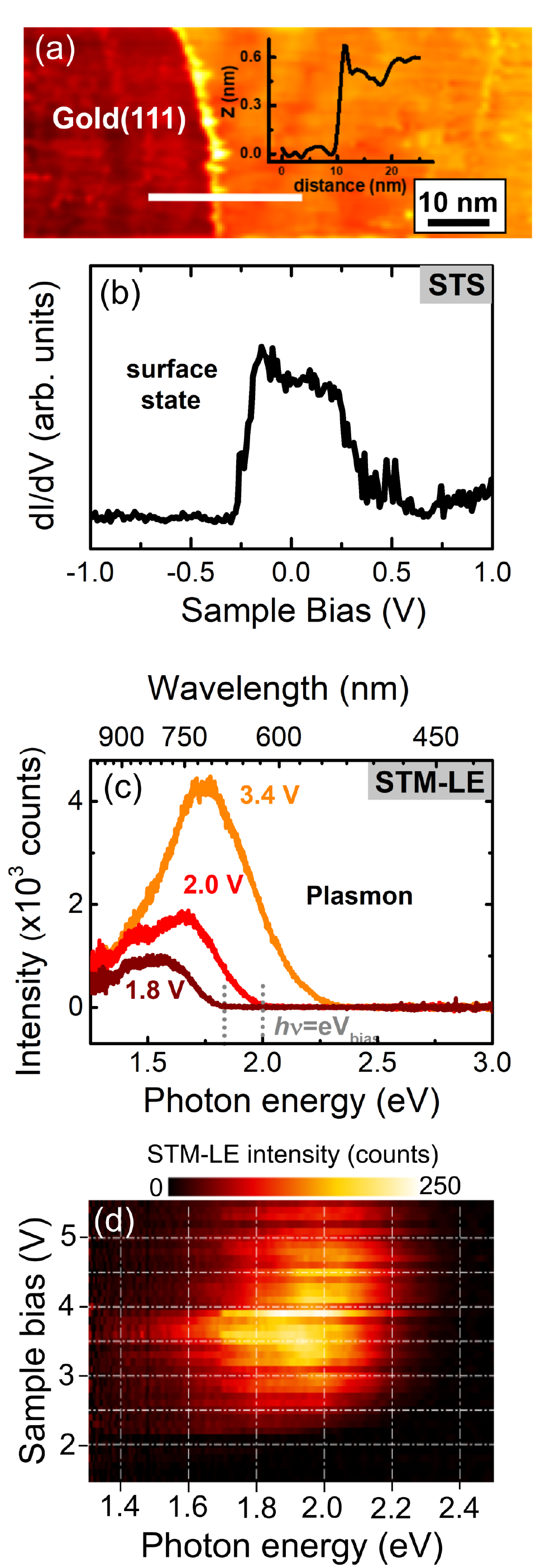}
\caption{\label{PlasmonGold} (a) STM image of the gold(111) surface (-1.5 nA, -210 mV), the inset figure shows the height profile along the surface step. (b) STS curve (80 mV, 0.8 kHz).) (a) and (b) were obtained at 80 K. (c) STM-LE obtained at 3 different bias values with a tunneling current of 10 nA and 30 s acquisition time. (d) Series of spectra as a function of the applied sample bias for a tunneling current of 10 nA and 5 seconds of exposure time. (c) and (d) were measured at 100 K.}
\end{figure}

Fig.~\ref{PlasmonGold} summarizes some topographic, spectroscopic and optical results obtained on a clean gold surface at LT. The sample consisted of a gold(111) thin film with large atomically flat areas.\cite{HEGNER1993} Before measurements, the sample was heated at 470 K under UHV conditions for removing undesired contaminants. The STM image and the STS curve presented in Fig.~\ref{PlasmonGold}(a) and (b), respectively, were acquired at 80 K, with the optical device retracted and thermal shields closed. The image shows a surface step with about 0.5 nm of height. The STS reveals the presence of a surface state between -0.5 V and the Fermi level, as expected for the gold surface.\cite{ROSCH2015,Andreev2004} When the thermal shields were opened and the mirror was inserted into the STM, the sample temperature increased up to 100 K and remained stable for several hours. 

Fig.~\ref{PlasmonGold}(c) shows three STM-LE spectra acquired at different sample biases and a constant tunneling current of 10 nA. The strong plasmonic emission was optimized after forming a gold tip by several indentations of a tungsten tip on the gold surface. The plasmonic spectra presented here are broad peaks whose final shape and maximum depend on the tip condition and geometry.\cite{ROSSEL2010, Aizpurua2000, martin2020} The inelastic tunneling process involved in the light emission mechanism is put in evidence by the quantum cut-off ($h\nu<$eV$_{\text{bias}}$) of the photons observed in the STM-LE spectra at low bias, from which it is shown that no photon can be emitted with an energy higher than the energy of tunneling electrons. \cite{martin2020} The luminescence spectra presented here were obtained using the setup illustrated in Fig.~\ref{OpticalExperiments}(b) in UHV and at a temperature of 100 K in the sample.

The dependence of the plasmonic light emission on the applied sample bias is better characterized by the voltage map shown in Fig~\ref{PlasmonGold}(d). In this figure, each line represents a spectrum acquired at the samples bias indicated by the vertical axis. The color represents the light intensity recorded by the CCD. The highest count-rate is observed with about 3.5 V. The decrease in emission at higher bias voltages is attributed to the increasing distance between the tip and the surface that happens to keep a constant current at varying sample bias voltages.~\cite{ROSSEL2010,Edelmann2018}

Changing the optical setup and sending all light into a PMT, as indicated in the setup of Fig.~\ref{OpticalExperiments}(d), we measured the quantum yield of the sample for some tip conditions. In this mode, only the viewport and one lens are between the mirror and the detector. Also, the quantum efficiency of PMTs is well documented allowing a straightforward calculation of the sample quantum yield. This is better than using the setup in ~\ref{OpticalExperiments}(b) which has more elements (optical fiber, spectrometer mirrors, and grating and the CCD) that add uncertainty to the quantum yield calculation. For an unreconstructed gold(111) surface, we observed quantum yields between 1.4x10$^{-4}$ to 1.4x10$^{-5}$ photons per electron. This compares well with literature values for systems using lens or optical fibers.~\cite{Watkins2007} This comparison is relevant as it, to a certain extent, validates the calculation of collection efficiency.

\subsection{\label{sec4.2}{STS and optical spectroscopy on Quantum Dots}}

\begin{figure}[H]
 \centering
\includegraphics [width=2.3 true in]
{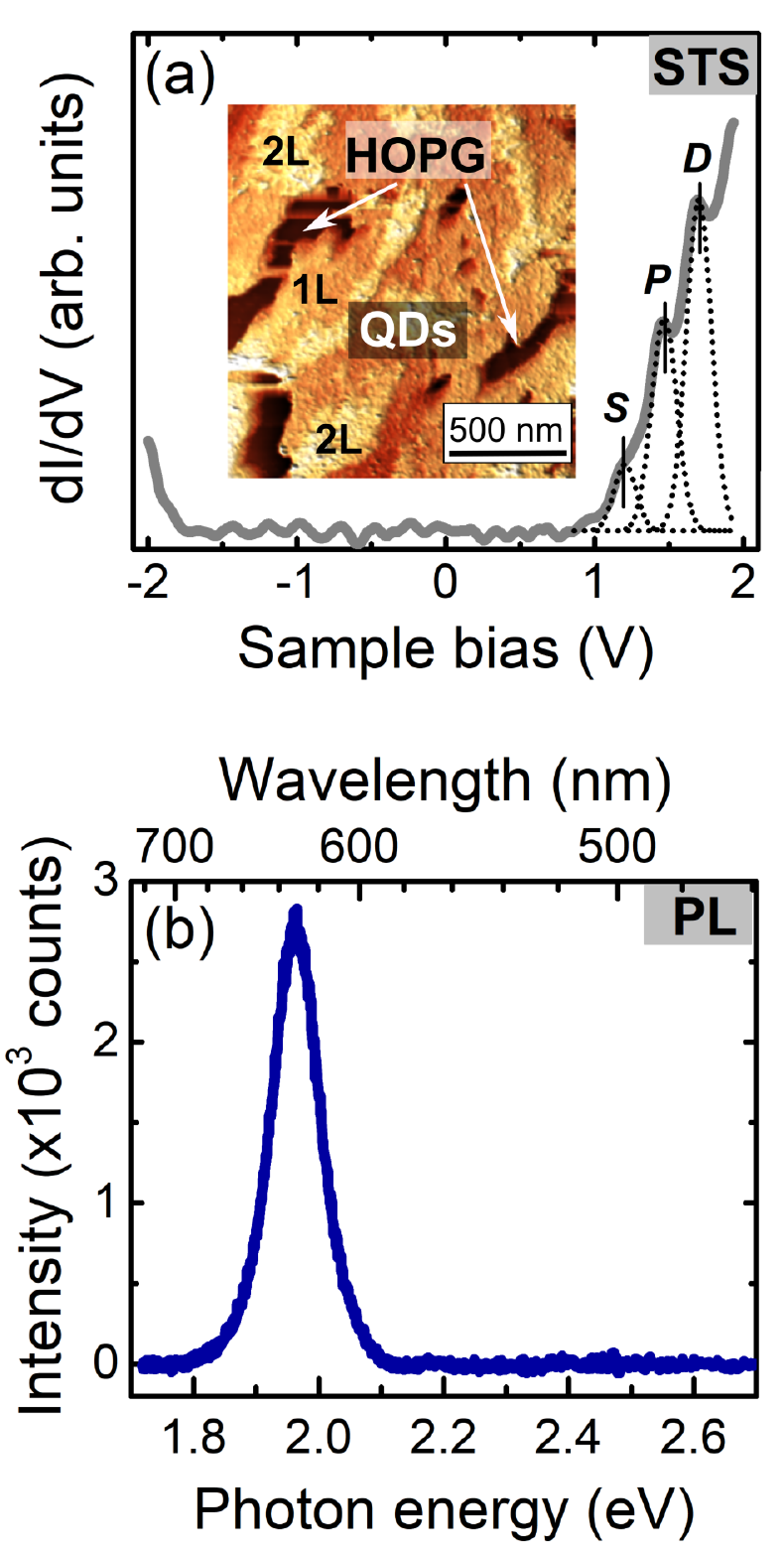}
\caption{\label{QDots} (a) STS curve (80 mV, 0.80 kHz) of 6 nm CdSe/ZnS QDots on HOPG in UHV conditions and at RT. Inset: Constant-current STM image showing one (1L) and two (2L) layers of Qdots on HOPG (2.5 V, 1.8 nA). (b) PL spectrum on the same region as the inset in (a) obtained on the sample sequentially and with the same conditions as the sample inside the STM.}
\end{figure}

Fig.~\ref{QDots} shows some results obtained on spherical core/shell CdSe/ZnS QDots with a diameter of about $~$6 nm and deposited on HOPG substrate. Differently from previous STM/STS results, these measurements were carried out with the parabolic mirror inserted inside the STM head and placed between the sample surface and the tip holder, as shown in Fig.~\ref{OpticalDevice}(c). The data were acquired at RT and under UHV. In general, the acquisition of STM images on QDs is always difficult because the capping ligands on the QDs surface produce instabilities in the tunneling current.\cite{Hummon2010} Therefore, to evaporate some ligands from the surface of the QDs and improve the STM imaging, before STM measurements the sample was annealed to 415 K for 4 hours in UHV. The inset in Fig.~\ref{QDots}(a) shows a large area STM image where regions with one and two layers of QDots are observed. The darkest regions in the image correspond to the HOPG surface. The STS curve presented in the figure was taken fixing the STM tip position on a region with one layer of QDots. This STS curve presents a zero dI/dV region of about 2.7 V and some confined electronic states in the conduction band which are associated with the atomlike energy levels S, P, and D, in perfect agreement with reported STS studies on individual QDots.\cite{ Diaconescu2013, Jdira2008, Overgaag2008, Liljeroth2006} Additionally, as explained above, the parabolic mirror can be used to inject light and measure the \emph{in situ} PL response of the sample. Fig.~\ref{QDots}(b) shows the PL spectrum obtained by injecting the laser light of 450 nm which shows an excitonic peak around 1.96 eV (632 nm).

The possibility of combining STM/STS with \emph{in situ} optical spectroscopy had been demonstrated to be an important approach to investigate the electronic structure and tunneling process in quantum-confined nanostructures.\cite{Katz2002,Millo2001} On the one hand, STS can give us information about both conduction band (CB) and valence band (VB) confined states, and, on the other hand, optical spectra give access to the allowed optical transitions between the states in the CB and VB. In the case of the results presented in Fig.~\ref{QDots}, we have that, in contrast to typical semiconducting surfaces like WSe$_2$, the electronic band gap of the QD is not given by the zero dI/dV gap region in the STS curve. The reason for that is that QDs are not in ohmic contact with the surface substrate, then a double barrier tunneling junction is formed between the STM tip, the QD, and the HOPG surface. This means that the bias applied to the substrate relative to the STM tip has to be distributed between two tunneling junctions.\cite{Hummon2010,Liljeroth2006} Therefore, the measured gap $\Delta V=2.7$ eV is always larger than the optical band gap ($E_{\text{opt}}$) of the QD by a factor $1/\eta$, where $\eta$ corresponds to the ratio between the potential drop in the tip-QD junction and the sample bias. The relation between the zero dI/dV gap with the optical gap is given by the following equation: $\eta\Delta V=E_{\text{opt}}+J_{e-h}$, where $J_{e-h}$ is the electron-hole Coulomb attraction energy.\cite{Liljeroth2006,Jdira2006,Bakkers2001} Considering $J_{e-h}\approx0.3$ eV as reported for Jdira, \emph{et al~}.\cite{Jdira2006}, and taking into account that the maximum of the PL peak in Fig.~\ref{QDots}(b) can be used as a measure of $E_{\text{opt}}$,\cite{Lee2019} it is possible to obtain that $\eta=0.8$. This value is very close to the theoretical value of $\eta=0.7$ calculated by Jdira, \emph{et al~}.\cite{Jdira2008} for one layer of CdSe/ZnS with 6 nm diameter QDs on HOPG substrate. The explanation given here illustrates how to use the data provided by the combination of tunneling and optical spectroscopies to test theoretical models of nanostructures. This also allows us to evaluate our setup and say that the mirror insertion around the tunneling junction does not affect the imaging and spectroscopic capability of the microscope.

\begin{figure}[h]
 \centering
\includegraphics [width=2.2 true in]
{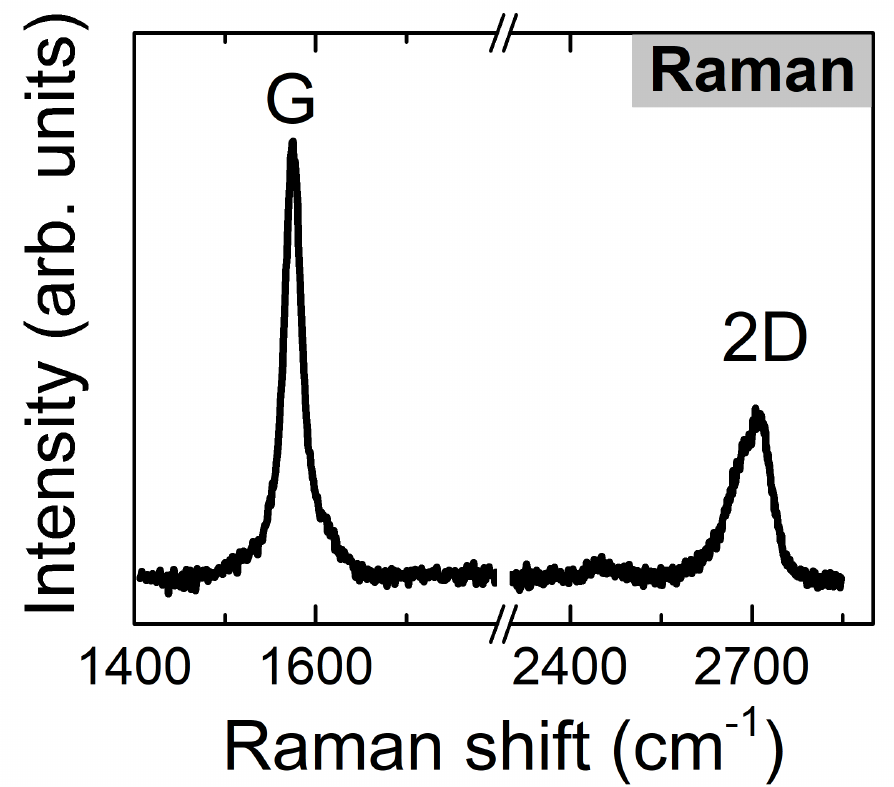}
\caption{\label{RamanHOPG} Raman spectrum of HOPG measured without the STM tip near the sample surface. The spectrum was acquired using a 532 nm laser diode (60 mW of power) and 1800 gr/mm grating. The acquisition time was 10 s and 4 acquisitions were averaged.}
\end{figure}

\begin{figure*}
\includegraphics[width=5.5 true in]
{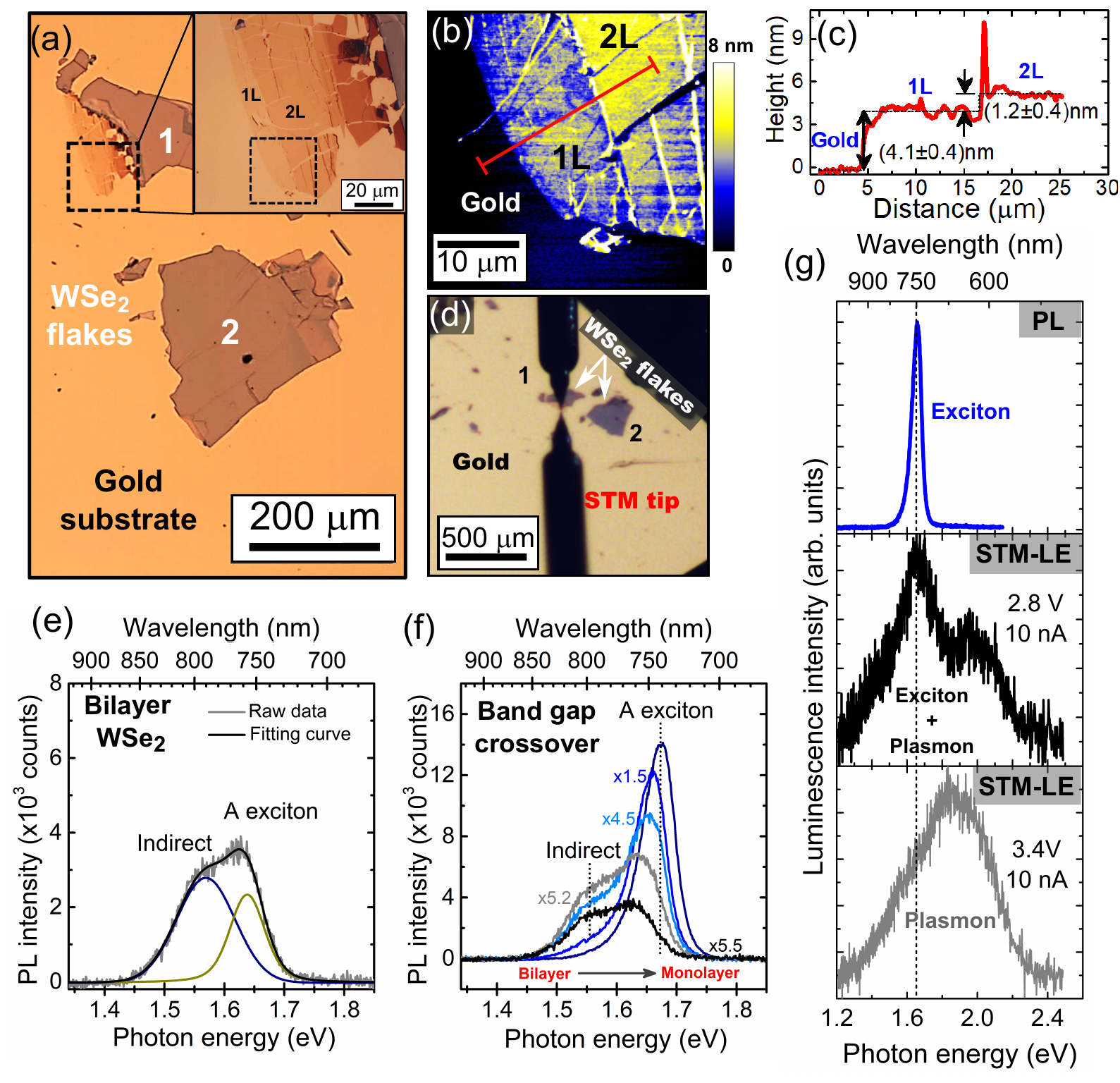}
\caption{\label{WSe2_AFM-STM-PL} 
 (a) Optical microscopy image to locate the flake and the region of interest as shown in the inset image. (b) AFM image of the region with monolayer (1L) and bilayer (2L) of WSe$_2$ indicated in (a). (c) Average height profile measured along the red line in (b) confirming the ML height. (d) Location of the flake inside the STM, the tip is approached on the flake of interest. (e) \emph{In situ} PL spectrum taken at room temperature on the bilayer, showing a decomposition in two peaks. (f) Band gap crossover observed in the transition from the bilayer to the monolayer region. (f) Comparison of the PL and STM-LE spectra acquired at RT and under UHV. The acquisition time for the STM-LE spectra was 10 s.}
\end{figure*}

\subsection{\label{sec4.3}{ \emph{In situ} Raman Spectroscopy on HOPG}}

Raman spectroscopy is a powerful technique for material characterization. Performing Raman spectroscopy requires an efficient injection and collection system due to the relatively weak signal yield and the proximity between the signal and the excitation laser. Performing Raman spectroscopy inside an LT UHV STM can be challenging due to the lack of space for the optics close to the tunnel junction.~\cite{Sheng2018} We performed Raman spectroscopy using a 532 nm laser diode and an 1800 gr/mm grating and keeping all other parameters identical to those used for usual STM-LE, that is, the same collecting optics and transmission into the spectrometer. The exception is the presence of a beam-splitter to inject the laser into the beam path, according to Figure ~\ref{OpticalExperiments}(c).

Fig.~\ref{RamanHOPG} shows an \emph{in situ} Raman spectrum measured in our setup. The result was obtained on a piece of HOPG, similar to the one used for Fig~\ref{STMPerformance}(a). The spectrum shows the characteristics G and 2D Raman bands at 1580 cm$^{-1}$ and 2700 cm$^{-1}$, respectively.\cite{zolyomi2011} This spectrum was acquired without the tungsten tip approached on the sample surface, being, hence, a regular (not tip-enhanced) Raman spectrum. For TERS experiments, gold or silver STM tips are required. \cite{Sheng2018,foti2018,Fang2016} 

\subsection{\label{sec4.4}{STM-LE on exfoliated monolayer WSe$_2$ }}

 \begin{figure*}
\includegraphics[width=5.6 true in]
{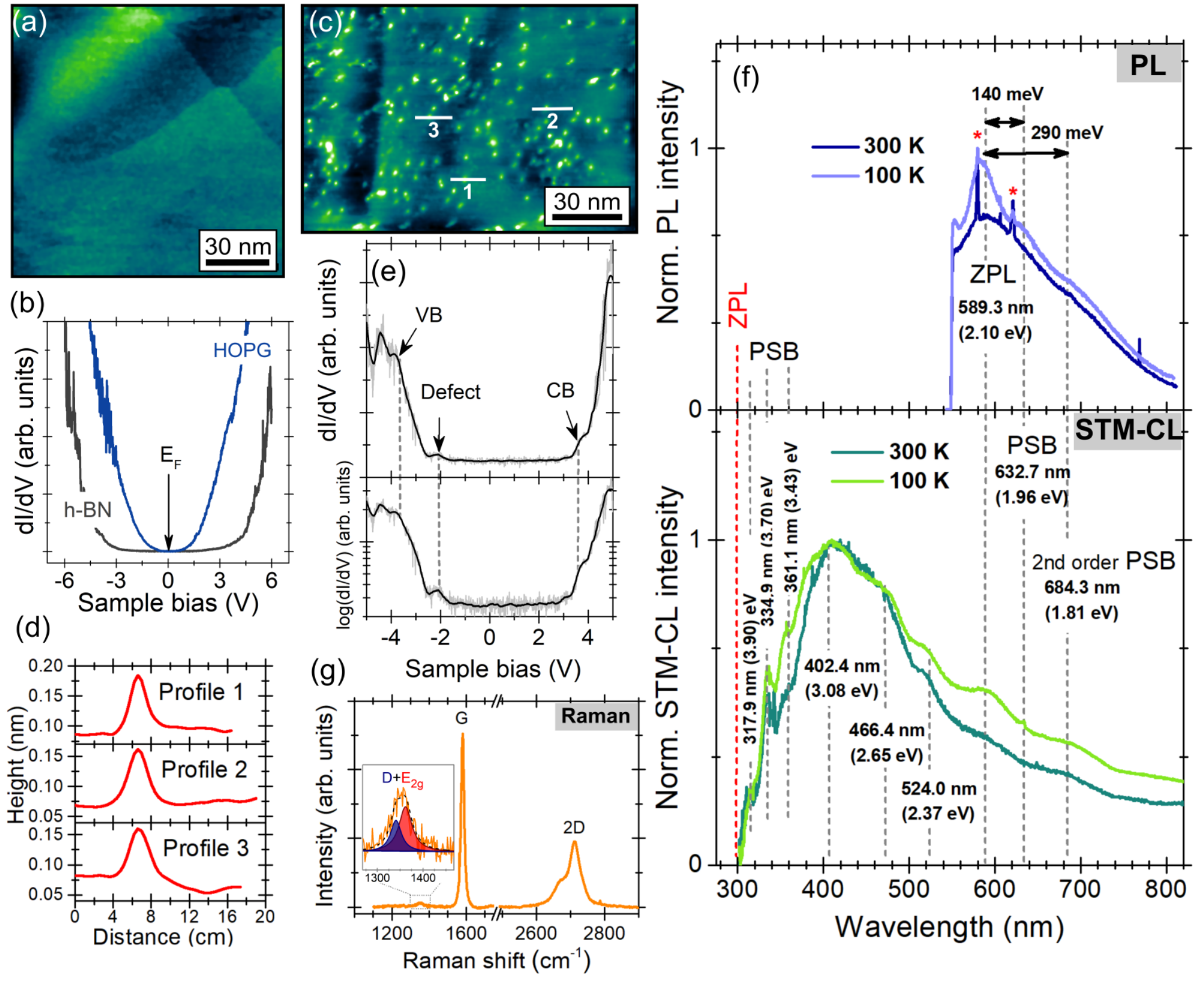}
\caption{\label{hBN} 
(a) STM images of the monolayer h-BN/HOPG surface free of defects (0.8 nA, 0.2 V, 80 K). (b) STS curves obtained in the region free of defects. (c) h-BN/HOPG surface with point defects (0.8 nA, 0.8 V, 80 K). (d) Height profiles of some defects. (e) STS curves obtained in the region with point defects. (f) \emph{In-situ} PL and STM-CL spectra in which Phonon Side-Bands (PSB) are observed near the Zero Phonon Line (ZPL) peaks. (g) \emph{In-situ} Raman spectrum at room temperature. Thes STS curves were recorded at 80 K using a bias modulation of 80 mV of amplitude and 800 Hz of frequency. Reproduced with permission from 2D Mater. 8 044001 (2021)\cite{Ricardo2021}. Copyright 2021 Institute of Physics.}
\end{figure*}

Observing micro-metric flakes of monolayers of 2D materials inside an STM requires the location of the object of interest under the STM tip and also, in our case, the alignment with the optical device. Fig.~\ref{WSe2_AFM-STM-PL} illustrates this process for a typical mechanically transferred WSe$_2$ micro flake containing a monolayer (in this case, we used gold as substrate). First, as shown in Fig.~\ref{WSe2_AFM-STM-PL}(a) and (b) the sample is characterized outside the STM using optical microscopy and Atomic Force Microscopy (AFM). This allows the identification of the appropriate flake among those on the substrate. Fig.~\ref{WSe2_AFM-STM-PL}(c) shows the expected height profile in as transferred TMDs flakes.\cite{Pena2020,Palleschi2020} Once inside the STM, a zoom-lens optical microscope helps finding the area of interest and approaching the tip accordingly, as shown in Fig.~\ref{WSe2_AFM-STM-PL}(d).

The light injection/collection device can be used to observe PL and the sample position can be adjusted according to the observed signal. Fig.~\ref{WSe2_AFM-STM-PL}(e) shows \emph{in situ} PL spectra obtained at RT that display both the A exciton and an indirect transition, as expected in WSe$_2$ bilayers.\cite{bagaev2019} The band gap crossover is shown in Fig.~\ref{WSe2_AFM-STM-PL}(f) where a transition from bilayer to monolayer is observed for several spectra recorded at different sample positions. The most intense spectrum shows only the A exciton peak, indicating the laser spot is hitting a monolayer. The spectra in Fig. ~\ref{WSe2_AFM-STM-PL}(e) and (f) were recorded with the STM tip retracted. 

Turning off the laser light, and approaching the tip until observing tunneling current, STM-LE signals could be recorded. In Fig. ~\ref{WSe2_AFM-STM-PL}(g) an \emph{in situ} PL spectra is shown for comparison with two STM-LE spectra. In one position, both the WSe$_2$ exciton peak and a plasmon due to the substrate are observed. In another region, only the plasmon peak is recorded. Plasmon emission is expected when the TMD collapses and enters in contact with the metallic substrate.\cite{Pena2020, Tumino2019, Bhanu2014} In some cases, however, the presence of a thin water layer explains the decoupling of the TMD and the metallic substrate which could quench the excitonic emission. The presence and quenching of the excitonic and plasmonic emissions are discussed in detail in ref. \cite{Pena2020}. Here, we would like just to point out that the STM-LE excitonic peak is aligned with the PL, while plasmon peaks shift in energy for different regions on the sample and tip conditions. 

 \subsection{\label{sec4.5}{STM-CL and emission due to defects in epitaxial monolayer h-BN}}
 
 h-BN is considered as a wide band gap semiconductor with many potential applications in deep ultraviolet sources and room temperature single photon emitters.\cite{Watabw2011,Jiang2014,tran_quantum_2016,toan2016} To unleash such applications, a better understanding of the fundamental electronic and optical properties of h-BN and the impact of structural defects is required. Fig.~\ref{hBN} summarizes some results obtained on monolayers h-BN epitaxially grown on HOPG. By using the experimental setup described in this work, h-BN monolayers were investigated by STM, STS, PL, Raman, and STM-CL measurements (see details in ref \cite{Ricardo2021}). STM images together with Raman spectroscopy provide us information about the morphology of the sample surface and the atomic structure of the system. From STS, PL, and STM-CL, it is possible to describe the electronic structure and optical transitions of the sample. Fig.~\ref{hBN}(a) shows an STM image of the sample surface free of defects. Since the h-BN and HOPG surfaces are morphological very similar, it is necessary to measure the STS curve of the scanned region to distinguish between them. The gray plot in Fig.~\ref{hBN}(b) corresponds to a typical STS curve obtained in the region shown in Fig.~\ref{hBN}(a). This dI/dV curve shows an electronic band gap of $\sim$6 eV, defined by the low differential conductance range from $\sim$-3 V to $\sim$+3 V confirming that the scanned region corresponds to an h-BN covered surface. In other regions with similar surface morphology, dI/dV curves with a zero band gap (like the blue curve in Fig.~\ref{hBN}(b)) expected to graphite are measured. The h-BN electronic band gap can be obtained by in-depth analysis of ensembles of STS data, as shown in ref\cite{Ricardo2021}. For a monolayer of h-BN on HOPG, an electronic band gap of $(6.8\pm0.2)$ eV was found. Regions with point defects were also identified in the sample. These defects appear as bright spots for positive sample bias, as can be seen in Fig.~\ref{hBN}(c). They have from 1 to 2 nm of diameter, according to the full width at half maximum in the height profiles, shown in Fig.~\ref{hBN}(d). STS measurements in regions with point defects, see Fig.~\ref{hBN}(e), indicate an acceptor level around -2 eV. 
 
 For measuring \emph{in situ} PL and Raman, the STM tip is retracted from the sample surface and the light of a green (532 nm) laser diode is injected into the STM in UHV. The \emph{in situ} PL spectra measured at 100 K and 300 K are shown in Fig.~\ref{hBN}(f). The PL spectra show a peak around 2.10 eV, which is clearly resolved at 100 K, and two shoulders around 1.96 and 1.81 eV. These emissions are typically associated with the zero-phonon line (ZPL) and two phonon replicas or phonon side bands (PSB) of carbon-related defects, respectively. \cite{mendelson_identifying_2021} The sharp peaks present in the PL spectra at 300 K, indicated with the red stars, correspond to the Raman response of the sample. The full Raman spectra can be seen in Fig.~\ref{hBN}(g), where the stronger peaks around 1580 and 2700 cm$^{-1}$ are the G and 2D Raman bands of HOPG, respectively, and the weak peak at 1350 cm$^{-1}$ coming from the D Raman mode of HOPG and the E$_{\text{2g}}$ mode of monolayer h-BN.
 
For STM-CL measurements, the STM tip was retracted about 150 nm from the sample surface and a high voltage between 150 and 180 V was applied to the sample, obtaining FE currents of 5-10 $\mu$A.\cite{Ricardo2021} It must be noted that usual CL experiments performed with fast electrons could only record CL signal for thicker h-BN samples with six or three monolayers of thickness.\cite{Schue2016,Hernandez2018} Here, by using STM-CL, it is possible to measure the CL response of a single monolayer. From the STM-CL spectra in Fig.~\ref{hBN}(f), it can be observed that the CL peaks are better resolved at 100 K than at 300 K. 
 The peaks related to carbon defects are observed at ZPL at 2.10 eV and a PSB at 1.81 eV. At low wavelengths, the spectra show some transitions at 3.90, 3.70, and 3.43 eV, which have been reported as being phonon replicas of a carbon-related defect emitting at 4.1 eV. The most intense peaks in the STM-CL spectra are identified at 3.08, 2.65, and 2.37 eV. These transitions have been previously reported for bulk h-BN and are normally associated with carbon impurities or nitrogen vacancy type-centers. \cite{hayee_revealing_2020,PhysRevB.100.155419,BERZINA2016131}

\section{\label{sec5} Luminescence experiments inside the STM in ambient conditions} 

Besides operating under UHV and at LT, the STM and light injection and collection device can be used in air (ambient conditions).\cite{Ricardo2021,DOAMARAL2021} Here are presented some results obtained in WSe$_2$ monolayers excited with the tunneling current when the STM operated in ambient conditions. 

\begin{figure}[H]
 \centering
\includegraphics [width=2.7true in]{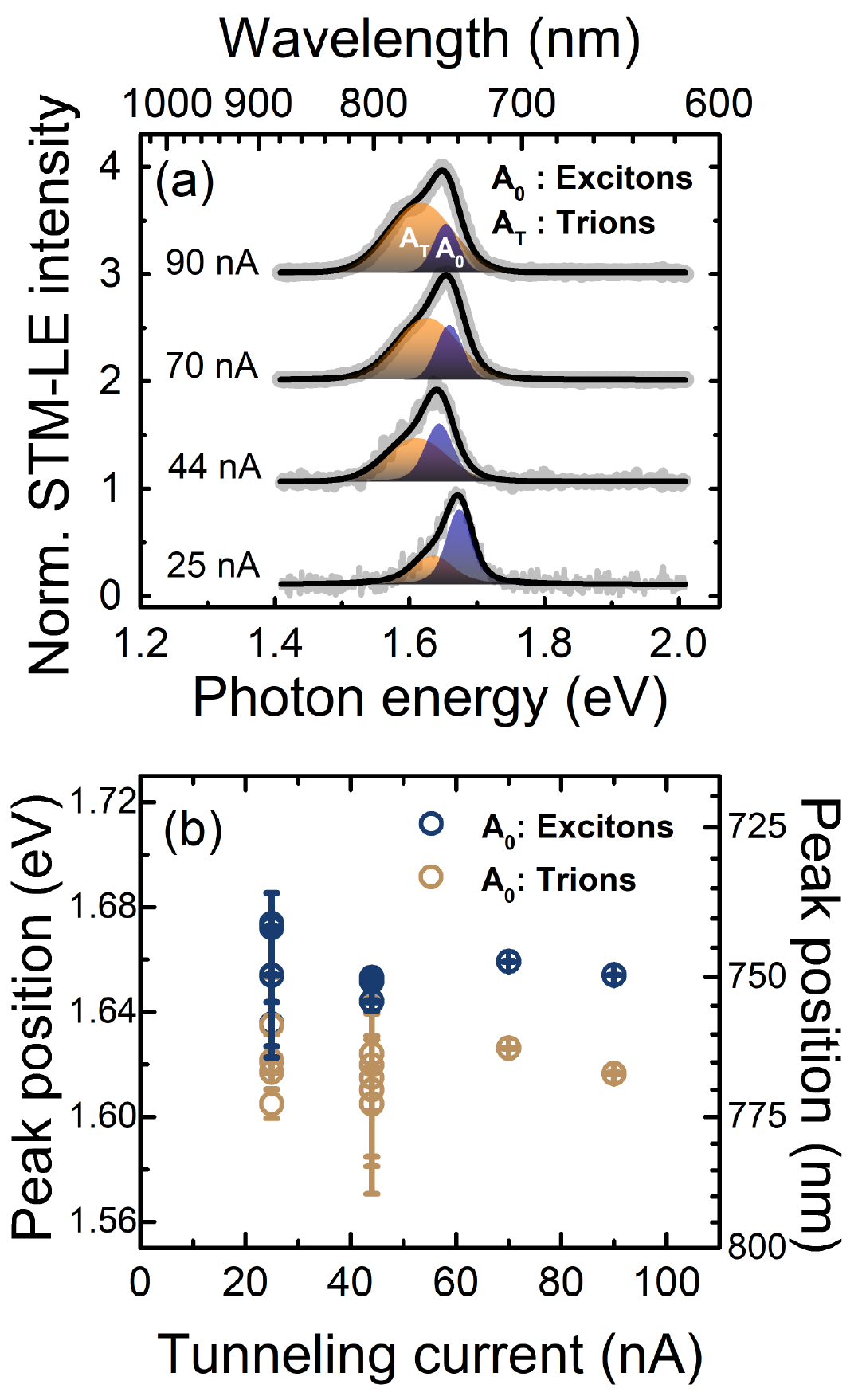}
\caption{\label{from_Nanoscale} (a) STM-LE spectra of the excitonic emission from WSe$_2$ recorded as a function of the tunneling current and for sample bias of 2.4 V. Neutral exciton and charged exciton (trions) contributions are indicated by the fit. (b) Emission energy as a function of the tunnel current for the charged and neutral exciton emission obtained from fitting emission spectra. For 25 and 44 nA, several measurements were made to gain some statistics. Reproduced with permission from Nanoscale, 12, 13460-13470(2020) \cite{Pena2020}. Copyright 2020 Royal Chemical Society. 
}
\end{figure}

A key point in STM-LE studies on semiconductors is the establishment of a partial coupling to a conductive support that allows the flow of the tunneling current but does not quench light emission due to all metallic states creating interface states inside the semiconductor band gap. This is particularly true for molecules and also holds for atomically thin 2D semiconductors.\cite{Kuhnke2017} Keeping the unintentional water layer that is present under mechanically transferred monolayers, it is possible to study monolayers of WSe$_2$ on top of metallic supports such as gold.\cite{Pena2020} Using this strategy, it was possible to observe STM induced excitonic light emission from WSe$_2$, as shown in Fig.~\ref{from_Nanoscale}. 

The light emission observed in Figure ~\ref{from_Nanoscale} has the same energy as recorded by PL, similarly as Figure ~\ref{WSe2_AFM-STM-PL}(f), and is basically independent of the applied sample bias voltage (see ref. ~\cite{Pena2020}), confirming its excitonic nature. TMDs such as WSe$_2$ are known to display charged exciton (trion) emission, a result of the high carrier density and strong Coulomb interaction. Using STM-LE, it was possible to, similarly as power dependent PL, observe the increase of the light emission associated with charged excitons with increasing tunnel current. Interestingly, the trion emission scales linearly with the tunnel current (see ref. \cite{Pena2020}). Moreover, despite the considerable change in trion to neutral exciton ratio seen in Fig.~\ref{from_Nanoscale}(a), the emission energy of these two transitions remains fairly constant, as shown in Fig.~\ref{from_Nanoscale}(b). These energies are indeed in close agreement with those obtained from PL. This indicates that the intense electric field under the STM tip does not necessarily cause emission energy shifts in the TMD, as one could expect.

\section{\label{6} Conclusions}

Some design criteria, the implementation, and some illustrative results of a new light detection scheme and an STM employing such system are presented. The design criteria show how the use of a parabolic mirror associated with the Brightness/\emph{Étendue} conservation theorems yields a high-performance light collection device that can be used in combination with an LT UHV STM to display 72$\%$ of collection efficiency, $50\%$ of total transmission efficiency with a 0.5 nm spectral resolution. Moreover, the limits regarding efficiency and its relation to the available field of view and spectral resolution were discussed. 

The results show that the performance of the STM was not affected by the presence of the light detection device or the modifications performed on the commercial STM system. Luminescence results including \emph{in situ} PL, \emph{In situ} Raman spectroscopy, STM-CL and STM-LE were shown for several systems, including h-BN, TMDs, QDots, metallic surfaces, and HOPG. This set of results demonstrates the potential of this light injection and collection device for studies of light emission and absorption in the context of an LT UHV STM. 

We expect these design criteria and implementation represents a solid base for the development of light injection and collection devices for other LT UHV STMs and also other types of microscopes. Indeed, several experiments involving light injection and/or collection around a tunnel junction or focused electron beam are possible with the presented device and such setup could contribute to several areas of photonics and plasmonics. 

\begin{acknowledgments}
This work was supported by the Fundação de Amparo à Pesquisa do Estado de São Paulo (FAPESP) Projects 14/23399-9 and 18/08543-7. I.D.B. acknowledges the financial support from the Brazilian Nanocarbon Institute of Science and Technology (INCT/Nanocarbono) and Brazilian Synchrotron Light Laboratory (LNLS).
\end{acknowledgments}

\section*{Author Declarations}

\subsection*{Conflict of Interest}
L.F.Z., R.J.F.R., and Y.A. have Patent No. BR102020015402-8 pending. L.F.Z., R.J.P.R., and Y.A. receive royalties in accordance with a technology transfer contract between Unicamp (represented by L.F.Z.) and RHK Technology, Inc.

\section*{DATA AVAILABILITY}
The data that support the findings of this study are available
from the corresponding author upon reasonable request.

\section*{REFERENCES}
\nocite{*}

\bibliography{aipsamp}

\end{document}